%% file: article.tex
\documentclass[11pt,a4paper]{article}

\usepackage{geometry,amsmath,amsfonts}
\usepackage{slashed}
\usepackage{epsfig}
\usepackage{hyperref}
\usepackage{latexsym}
\usepackage{graphicx}
\usepackage{amssymb}
\usepackage{verbatim}
\usepackage{bm}
\usepackage{dsfont}
\usepackage[footnotesize]{caption}
\usepackage{cancel}
\usepackage{cite}
\usepackage{graphicx}

\usepackage{epsfig}

\usepackage{multirow}
\usepackage{tikz}
\usepackage{subfig}
\usepackage{color}
\usepackage{multirow}
\usepackage{color}
\usepackage{rotating}
\usepackage{ifthen}

\begin{document}
\thispagestyle{empty}
\setcounter{page}{0}

\vskip 1.5cm

\begin{center}
{\LARGE\bf Primordial gravitational waves for universality classes of pseudoscalar inflation.}

\vskip 2cm

{\large  Valerie~Domcke\footnote{valerie.domcke@apc.univ-paris7.fr}, Mauro Pieroni\footnote{mpieroni@apc.univ-paris7.fr}, Pierre Bin\'etruy\footnote{Member of the Institute universitaire de France.}}\\[3mm]
{\it{
 AstroParticule et Cosmologie (APC)/Paris Centre for Cosmological Physics, Universit\'e Paris Diderot, CNRS, CEA, Observatoire de Paris, Sorbonne Paris Cit\'e University.  \\

}}
\end{center}

\vskip 1cm

\begin{abstract}
Current bounds from the polarization of the CMB predict the scale-invariant gravitational wave (GW) background of inflation to be out of reach for upcoming GW interferometers. This prospect dramatically changes if the inflaton is a pseudoscalar, in which case its generic coupling to any abelian gauge field provides a new source of GWs, directly related to the dynamics of inflation. This opens up new ways of probing the scalar potential responsible for cosmic inflation. Dividing inflation models into universality classes, we analyze the possible observational signatures. One of the most promising scenarios is Starobinsky inflation, which may lead to observational signatures both in direct GW detection as well as in upcoming CMB detectors. In this case, the complementarity between the CMB and direct GW detection, as well as the possibility of a multi-frequency analysis with upcoming ground and space based GW interferometers, may provide a first clue to the microphysics of inflation.
\end{abstract}

\newpage

\tableofcontents

\section{Introduction}
\input{introduction}

\section{Review of the mechanism. \label{sec:review}}
\input{mechanism_0210}

\section{General analytical results. \label{sec:analytical}}
\input{analytical_0210}

\section{Numerical results for explicit models. \label{sec:models}}
\input{models}

\section{Further experimental signatures and bounds. \label{sec:bounds}}
\input{experimental}

\section{Discussion. \label{sec:discussion}}

\input{discussion}

\section{Conclusion and Outlook. \label{sec:conclusion}}
\input{conclusions}

\appendix
\section{Analytical estimates for models with $p \neq 2$. \label{sec:Appendix}}

\input{appendix}

\vspace{1cm}
\subsubsection*{Acknowledgements}

We thank Chiara Caprini, Daniel Figueroa, Jonathan Ganc, Daan Meerburg, Joel Meyers, Marco Peloso, Antoine Petiteau, Angelo Ricciardone, Ryo Saito, Martin Sloth, Kai Schmitz and  Lorenzo Sorbo for very helpful discussions. We also thank the eLISA cosmology working group for stimulating discussions and Stefan Rodrigues Sandner for a careful reading of the manuscript. We acknowledge the financial support of the UnivEarthS Labex program at Sorbonne Paris Cit\'e (ANR-10-LABX-0023 and ANR-11-IDEX-0005-02) and the Paris Centre for Cosmological Physics.

\input{refs}

\end{document}

%% file: introduction.tex
The recent first detection of gravitational waves (GWs) by the LIGO/VIRGO collaboration~\cite{Abbott:2016blz} has opened up a fascinating new window to the very early universe. Unlike any other messenger, GWs can travel freely through the Universe, carrying information on times as early as cosmic inflation. They are thus a unique and ultimate probe of any model of cosmic inflation. Current indirect bounds on primordial GWs, obtained from bounds on the Cosmic Microwave Background (CMB) polarization, indicate that the nearly scale-invariant spectrum of the vacuum fluctuations during standard slow-roll inflation lies below the range of current and upcoming direct GW detectors. This picture may however change dramatically once interactions of the inflaton with other particles are taken into account~\cite{Cook:2011hg}. \\

\noindent To this end, we consider the generic coupling of a \emph{pseudoscalar} inflaton $\phi$ to the field strength tensor $F^{\mu \nu}$ of any gauge groups of the theory,
\begin{equation}
{\cal L}_\text{int} \sim \phi F_{\mu \nu} \tilde{F}^{\mu \nu} \,.
\label{eq:Lint}
\end{equation}
Such pseudoscalar  (or `axionic') flat directions, suitable for inflation, may be expected to be abundant at the high energy scales of cosmic inflation. The coupling to gauge fields, even to the SM gauge fields, is then allowed by all symmetries of the theory and unavoidable from the point of view of effective field theory.\\

\noindent The presence of the interaction term~\eqref{eq:Lint} has a dramatic impact on the predictions of a given inflation model, classified by its scalar potential $V(\phi)$. This coupling to the gauge fields leads to a tachyonic instability for one of the gauge field modes and consequently to a non-perturbative production of the gauge field during inflation~\cite{Turner:1987bw, Garretson:1992vt, Anber:2006xt}. This in turn back-reacts on the dynamics of the inflaton~\cite{Anber:2009ua,Barnaby:2011qe,Barnaby:2011vw}, the usual Hubble friction term is superseded by a new friction term sourced by the gauge fields. Moreover, the gauge fields provide an additional source of tensor perturbations, leading to a potentially large enhancement of the resulting GW signal~\cite{Anber:2012du}.\\

\noindent We analyze the consequences of this for the predictions of cosmic inflation. So far, in this context the focus has mainly been on models of Chaotic inflation, see e.g.~\cite{Barnaby:2010vf,Barnaby:2011vw,Linde:2012bt}, which have served as useful toy-models to illustrate many of the relevant effects. In this paper, we take a broader approach, identifying universal features shared by all pseudoscalar single-field slow-roll inflation models but also pointing out how a coupling as in Eq.~\eqref{eq:Lint} can be an intriguing possibility to distinguish different inflation models. To this end, we group inflation models in terms of universality classes~\cite{Mukhanov:2013tua, Binetruy:2014zya} and analyze their parameter dependencies both analytically and numerically. In particular many of our analytical results are derived with basically no assumptions on the underlying inflation model, and may thus easily be adapted to specific models of interest. Our main results are on the CMB observables and the GW spectrum, but we also consider constraints from primordial black hole formation, the generation of primordial magnetic fields and Big Bang Nucleosynthesis (BBN) constraints. \\

\noindent Among the different universality classes, we find Starobinsky inflation~\cite{Starobinsky:1980te}\footnote{Pseudoscalar fields effectively describing the Starobinsky model of inflation can easily be described in the context of supergravity by employing a shift-symmetry in the K\"ahler potential~\cite{Kawasaki:2000yn} (see for example~\cite{Dall'Agata:2014oka,Kallosh:2010ug,Kallosh:2010xz}). Such models may be difficult to obtain from string theory because the presence of a coupling such as \eqref{eq:Lint} is associated with the presence of a pseudo-anomalous $U(1)$. But obtaining an ultraviolet completion of the models that we consider here goes beyond the scope of this paper.} to be the most promising from the point of view of possible observations - remarkably just the model which is one of the most favoured by the recent results of the Planck satellite~\cite{Ade:2015lrj}. In this case, the complementarity between direct and indirect searches of GWs at various frequencies, together with further CMB observations, will cover a significant part of the parameter space in the near future. Key experiments in the next decade will be ground- and space based GW experiments, as well as 4th generation CMB experiments.\\ 

\noindent This paper is organized as follows. We begin with a brief review  of pseudoscalar single-field slow-roll inflation in the presence of the interaction~\eqref{eq:Lint} in Sec.~\ref{sec:review}. This sets the stage to derive model-independent, analytical expressions for the most important quantities in Sec.~\ref{sec:analytical}. In Sec.~\ref{sec:models} we specify the relevant universality classes of inflation and proceed with a full numerical analysis of representative models, demonstrating also the validity of our analytical expressions. Some of the details are relegated to the appendix. Finally, we discuss further implications of our results in Sec.~\ref{sec:discussion}, considering also a possible embedding into a broader picture, both from the point of view of particle physics model building as well as for early universe cosmology. Special attention is given to the Starobinsky model, for which we present a full parameter scan investigating the observational prospects. We conclude in Sec.~\ref{sec:conclusion}.

%% file: mechanism_0210.tex
The production of gauge field quanta during inflaton has been studied in Refs.~\cite{Turner:1987bw, Garretson:1992vt, Anber:2006xt}. The resulting GW spectrum was analyzed in~\cite{Anber:2012du}, with the backreaction of the gauge fields on the inflationary dynamics considered in~\cite{Anber:2009ua,Barnaby:2011qe,Barnaby:2011vw}. Recently, the diverse phenomenology of this setup has received a lot of attention, including the study of primordial black hole (PBH) formation and CMB non gaussianities~\cite{Barnaby:2010vf,Linde:2012bt,Meerburg:2012id}. In this section we review the dynamics of the background fields and their fluctuations, setting the stage for our analysis of inflation models in Secs.~\ref{sec:analytical} and \ref{sec:models}.

\subsection{Background field equations}

Our starting point is a pseudoscalar inflaton $\phi$, coupled to ${\cal N}$ $U(1)$ gauge fields $A_\mu^a$, 
\begin{equation}
{\cal L}=  -\frac{1}{2} \partial_\mu \phi \partial^\mu \phi  - \frac{1}{4} F^a_{\mu \nu} F_a^{\mu \nu} - V(\phi) - \frac{\alpha^a}{4 \Lambda} \phi F^a_{\mu \nu} \tilde F_a^{\mu \nu} \,.
\end{equation}
Here $F^a_{\mu \nu}$  ($\tilde F_a^{\mu \nu}$) is the  (dual) field strength tensor, $\Lambda$ is the mass scale suppressing higher-dimensional operators of the theory and $\alpha$ parametrizes the strength of the inflaton - gauge field coupling. For simplicity, we will consider $\alpha^a = \alpha$ for all $a = \{1,2,..{\cal N}\}$ in the following.
The resulting background equations for $\phi(t)$ and $A^a(t,x)$ are
\begin{align}
\label{eq:eq_motion}
\ddot \phi + 3 H \dot \phi + \frac{\partial V}{\partial \phi} & = \frac{\alpha}{\Lambda} \langle \vec E \vec B \rangle \,.\\
\frac{d^2}{d \tau^2}\vec A^a - \nabla^2 \vec A^a - \frac{\alpha}{\Lambda} \frac{d \phi}{d \tau} \nabla \times \vec A^a & = 0 \,, \label{eq:eq_motionA}
\end{align}
where dots are used to denote derivatives with respect to cosmic time $t$, whereas $\tau$ denotes the conformal time. The Friedmann equation reads:
\begin{equation}
\label{eq:friedmann}
3 H^2 = \frac{1}{2} \dot \phi^2 + V + \frac{1}{2} \langle \vec E^2 + \vec B^2 \rangle \,.
\end{equation}
Assuming $\dot \phi$ is a slowly varying function in time, we can solve the equation for $\vec A$ analytically. The Fourier modes of $\vec A$ must satisfy:
\begin{equation}
\frac{ d^2 A^a_\pm(\tau, k)}{d \tau^2} + \left( k^2 \pm \frac{\alpha k \dot \phi}{\Lambda H \tau } \right)A^a_\pm(\tau, k) = 0 \,.
\label{eq:eq_motionA2}
\end{equation}
Here the subscript $\pm$ refers to the two helicity modes of the massless gauge field ($\vec{A}^{a} = \vec{e}_{\pm} A_{\pm}^a \exp(i \vec{k}\vec{x})$). The corresponding helicity vectors $\vec e_\pm(\vec k)$  satisfy $\vec k \times \vec e_\pm = \mp i k \vec e_{\mp}$, turning the cross-product in Eq.~\eqref{eq:eq_motionA} (arising in turn from the antisymmetric $\epsilon$-tensor in $\tilde F_{\mu \nu}$) into the $\pm$ in Eq.~\eqref{eq:eq_motionA2}. This leads to a tachyonic instability in the $A_+$ mode (for $\dot \phi < 0$) and hence to an exponential growth of one of the two helicity modes of the vector field,
\begin{equation}
A_+^a \simeq \frac{1}{\sqrt{2k}} \left( \frac{k}{2 \xi a H}\right)^{1/4} e^{ \pi \xi - 2 \sqrt{2 \xi k/(a H)}} \,,
\end{equation}
where we have defined
\begin{equation}
\xi \equiv \frac{\alpha |\dot \phi|}{2 \Lambda H} \,.
\label{eq:xi}
\end{equation}
W.l.o.g., let us assume that $ \phi > 0, V'(\phi)>0, \dot \phi < 0$. The strong gauge field production modifies the slow-roll equation of motion and the Friedmann equation through\footnote{More precisely, and relevant for small $\xi$ \cite{Anber:2009ua}:
\begin{equation}
\langle \vec E \vec B \rangle \simeq  \frac{H^4}{\xi^4} e^{2 \pi \xi} \frac{1}{2^{21} \pi^2} \int_0^{8 \xi}
  x^7 e^{-x} dx \,.
\end{equation}
}
\begin{equation}
\langle \vec E \vec B \rangle \simeq {\cal N} \cdot \,  2.4 \cdot 10^{-4} \frac{H^4}{\xi^4} e^{2 \pi \xi} \,, \quad \frac{1}{2} \langle \vec E^2 + \vec B^2 \rangle  \simeq {\cal N} \cdot \,  1.4 \cdot 10^{-4} \frac{H^4}{\xi^3} e^{2 \pi \xi} \,.
\end{equation}
Typically the effect in the Friedmann equation is small. However, in the slow-roll equation for the inflaton, this introduces an additional friction term which can slow down inflation significantly as $\xi \sim |\dot \phi|/H$ increases towards the end of inflation. Inflation then extends for $\Delta N_*$ additional e-folds, implying that for a given scalar potential, the point where the CMB probes this scalar potential is shifted. Note that this does not change the total amount of e-folds ($N_\text{CMB} \simeq 60$) after the CMB scales exited the horizon, but these are now divided among $N_*$ efolds of standard inflation governed by the Hubble friction and $\Delta N_*$ e-folds of inflation governed by the gauge field induced friction, see also Fig.~\ref{fig:phi_xi_schematic}.

\subsection{Scalar and tensor perturbations}
Expressing the pseudoscalar field as $\phi(t,x) = \phi(t) + \delta \phi(x,t)$ the equation of motion for the scalar fluctuations reads:
\begin{equation}
\ddot{ \delta \phi} + 3  H \dot{ \delta \phi}+ (- \nabla^2 + V''(\phi)) \delta\phi = - \frac{\alpha}{\Lambda} \delta[\vec E^a \vec B^a] \,,
\label{eq:scalfluc}
\end{equation}
with
\begin{equation}
\delta[\vec E^a \vec B^a] = [\vec E^a \vec B^a - \langle \vec E^a \vec B^a \rangle]_{\delta \phi = 0} + \frac{\partial \langle \vec E^a \vec B^a \rangle}{\partial \dot \phi}\dot{\delta \phi} \,.
\end{equation}
In the regime of strong gauge fields, Eq.~\eqref{eq:scalfluc} can be solved approximately by considering only the second term on the lefthand side, as well as the gauge field terms on the righthand side, leading to~\cite{Linde:2012bt}:
\begin{equation}
\Delta_s^2(k) = \Delta_s^2(k)_\text{vac} + \Delta_s^2(k)_\text{gauge} = \left(\frac{H^2}{2 \pi |\dot \phi|}\right)^2 + \left( \frac{\alpha \langle \vec E \vec B \rangle/ \sqrt{\cal N}}{3 \Lambda \beta H \dot \phi} \right)^2 ,
\label{eq:scalar}
\end{equation}
with
\begin{equation}
\beta \equiv 1 - 2 \pi \xi \frac{\alpha \langle \vec E \vec B \rangle}{\Lambda 3 H \dot \phi}\,.
\end{equation} 
At large scales, gauge contributions are small and the spectrum approaches the scale-invariant spectrum of the standard vacuum fluctuations during inflation. At small scales, the gauge contributions dominate and the spectrum is given by:
\begin{equation}
\Delta_s^2(k ) \simeq \frac{1}{{\cal N} (2 \pi \xi)^2} \,.
\label{eq:scalar2}
\end{equation}
Note that the gauge fields affect the scalar spectrum in twofold way: by modifying the background slow-roll equation of motion and by modifying the equation of motion for the fluctuations directly. A more refined calculation of the solution to Eq.~\eqref{eq:scalfluc} can be found in \cite{Barnaby:2011vw}, with which the estimate above agrees up to an order one factor. \\

 \noindent
The tensor fluctuations
 are governed by the linearized Einstein equation~\cite{Maggiore:1900zz}:
 \begin{equation}
\frac{d^2 h_{ij}}{d \tau^2} + 2 \frac{d \ln a}{d \tau} \frac{d h_{ij}}{d \tau} - \Delta h_{ij} = \frac{2}{M_P^2} \Pi_{ij}^{ab} T_{ab} \,.
\label{eq:greens}
 \end{equation}
 Here  $\Pi^{ij}_{ab}$ is the transverse, traceless projector and $T_{ab}$ is the energy momentum tensor sourcing the gravitational waves. Eq.~\eqref{eq:greens} is solved by employing the Greens function $G_k(\tau, \tau_1)$ for the corresponding homogeneous differential equation,
 \begin{equation}
h_{ij}(k,\tau) = \frac{2}{M_P^2}\int d\tau_1 G_k(\tau, \tau_1) \Pi_{ij}^{ab}(k) T_{ab}(k, \tau_1)\,.
 \end{equation}
Inserting the background solution for the gauge fields, the amplitude of the tensor perturbations is given by (see e.g.\ \cite{Barnaby:2011qe, Barnaby:2011vw}):
\begin{equation}
\Omega_{GW} = \frac{1}{12} \Omega_{R,0} \left( \frac{H}{ \pi M_P} \right)^2 (1 + 4.3 \cdot 10^{-7} {\cal N} \frac{H^2}{M_P^2 \xi^6} e^{4 \pi \xi})\,,
\label{eq:OmegaGW}
\end{equation}
with $\Omega_{R,0} = 8.6 \cdot 10^{-5}$ denoting the radiation energy density today and $M_P = 2.4 \cdot 10^{18}$~GeV denoting the reduced Planck mass, which in the following expressions we will set to unity. Here the first term in the bracket is the usual vacuum contribution from inflation, whereas the second term is sourced by the contribution of the gauge fields to the anisotropic stress energy tensor. 
\\

\noindent
Finally, to depict the power spectra as a function of frequency, we employ:
\begin{equation}
N = N_\text{CMB} + \ln \frac{k_\text{CMB}}{0.002 \text{ Mpc}^{-1}} - 44.9 - \ln\frac{f}{10^2 \text{ Hz}},
\label{eq:Nf}
\end{equation}
with  $k_\text{CMB} = 0.002 \text{ Mpc}^{-1}$ and $N_\text{CMB} \sim 50 - 60$. In this convention, the number of e-folds $N$ decreases during inflation, reaching $N=0$ at the end of inflation.

%% file: analytical_0210.tex
In the equations of Sec.~\ref{sec:review}, the inflaton potential $V(\phi)$ was not further specified. Let us now turn to this point in more detail. We will follow the classification of inflation models of Ref.~\cite{Mukhanov:2013tua, Binetruy:2014zya}, which covers the vast majority of single-field slow-roll inflation models and is based on expressing the first slow-roll parameter $\epsilon$ as 
\begin{equation}
\label{eq:Nparameterization}
\epsilon_\phi \simeq \epsilon_V \simeq  \frac{\beta_p}{N^p}  + {\cal O}(1/N^{p+1})\,,
\end{equation}
where $\beta_p$ is a positive constant and $p$ is an integer and
\begin{equation}
\label{eq:epsilon_HV}
\epsilon_\phi = \frac{\dot \phi^2}{2 H^2}\,, \quad \epsilon_V = \frac{1}{2} \left( \frac{V'}{V} \right)^2 \,.
\end{equation}
The parametrization of the slow-roll parameters in powers of $1/N$ is a natural way to parametrize the observed smallness of the slow-roll parameters at the CMB-scales while accounting for an increase over the course of inflation, required to end inflation~\cite{Mukhanov:2013tua, Binetruy:2014zya,Roest:2013fha}. In the following, we will focus on inflation models obeying Eq.~\eqref{eq:Nparameterization} but otherwise keep the function $V(\phi)$ completely general. This enables a very general analysis of pseudoscalar inflation, with all the dependence on the underlying inflation model encoded in the parameters $p$ and $\beta_p$.\\

\noindent
The parametrization of Eq.~\eqref{eq:Nparameterization} is particularly convenient for analysis since the parameter $\xi$ (cf.\ Eq.~\eqref{eq:xi}), governing the strength of the gauge field production, can be expressed as
\begin{equation}
\xi \propto \sqrt{\epsilon_\phi} \,.
\label{eq:xigrowth}
\end{equation}
As can be seen from Eqs.~\eqref{eq:scalar}, \eqref{eq:scalar2} and \eqref{eq:OmegaGW}, the evolution of this parameter is crucial for the background dynamics as well as for the scalar and tensor spectrum. As long as the gauge fields are sub-dominant, $\xi$ grows as $1/N^{p/2}$. Once the gauge fields become significant, their backreaction effects the background evolution leading to a slower growth in $\xi$. This evolution of $\xi$ is reflected in the scalar and tensor power spectrum.\\

\begin{figure}
\begin{tikzpicture}
\node at (-4,0) { \includegraphics[width=0.45 \textwidth]{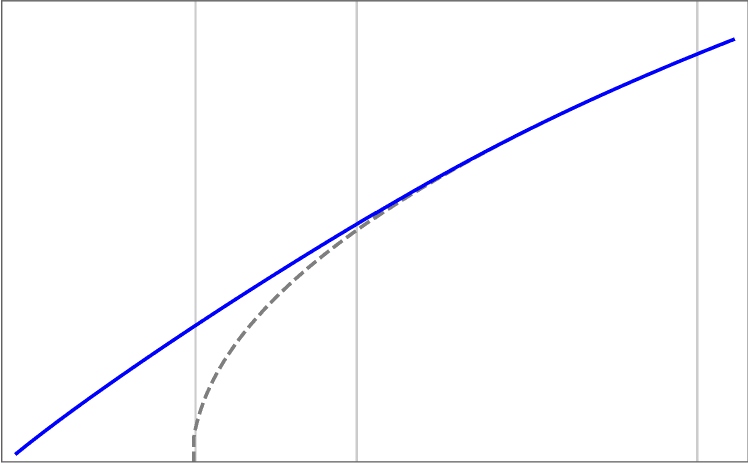}};
\node at (-4,-3) {$N$};
\draw[->] (-3.5, -3) -- (-2.5, -3);
\node at (-8, 0) {$\phi$};
\draw[->] (-8, 0.5) -- (-8, 1.5);
\node at (-5.8, -2.4) {\small $\Delta N_*$};
\node at (-4.2, -2.4) {\small $N_2$};
\node at (-1.0, -2.4) {\small $N_\text{CMB}$};
\node at (-7.5, -2.4) {\small 0};
\draw[<->] (-5.7, -1.4) -- (-1, -1.4);
\node at (-3, -1) {\small $N_*$};
\draw[<->] (-7.5, 1) -- (-4.2,1);
\node at (-6,1.5) {\small (C)};
\draw[<->] (-4.2,1) -- (-1,1);
\node at (-3,1.5) {\small (A,B)};
\node at (-6.7,-1) {\textcolor{blue}{\small $\alpha \neq 0$}};
\node at (-4.5, -1.0) {\textcolor{gray}{\small $\alpha = 0$}};
\node at (4,0) { \includegraphics[width=0.45 \textwidth]{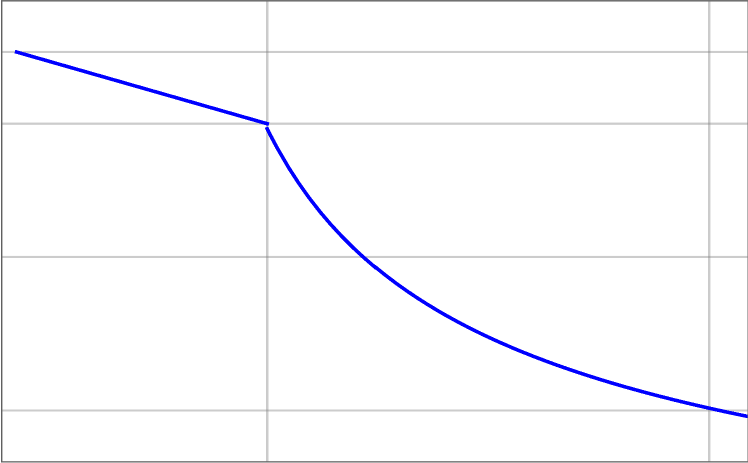}};
\node at (4, -3) {$N$};
\draw[->] (4.5, -3) -- (5.5, -3);
\node at (0.2, 0) {$\xi$};
\draw[->] (0.2, 0.5) -- (0.2, 1.5);
\node at (3.25, -1.9) {\small $N_2$};
\node at (6.5, -1.9) {\small $N_\text{CMB}$};
\node at (1, -1.5) {\small $\xi_\text{CMB}$};
\node at (0.8, 0.8) {\small $\xi_2$};
\node at (0.8, -0.45) {\small $\xi_1$};
\node at (1, 1.9) {\small $\xi_\text{max}$};
\node at (5.3, 0.3) {\small $\xi \propto (N - \Delta N_*)^{-p/2}$};
\end{tikzpicture}
\caption{Schematic view of the evolution of the inflaton field $\phi$ (left panel) and the parameter $\xi$ controlling the influence of the gauge fields  (right panel) as a function of the number of e-folds of inflation.}
\label{fig:phi_xi_schematic}
\end{figure}
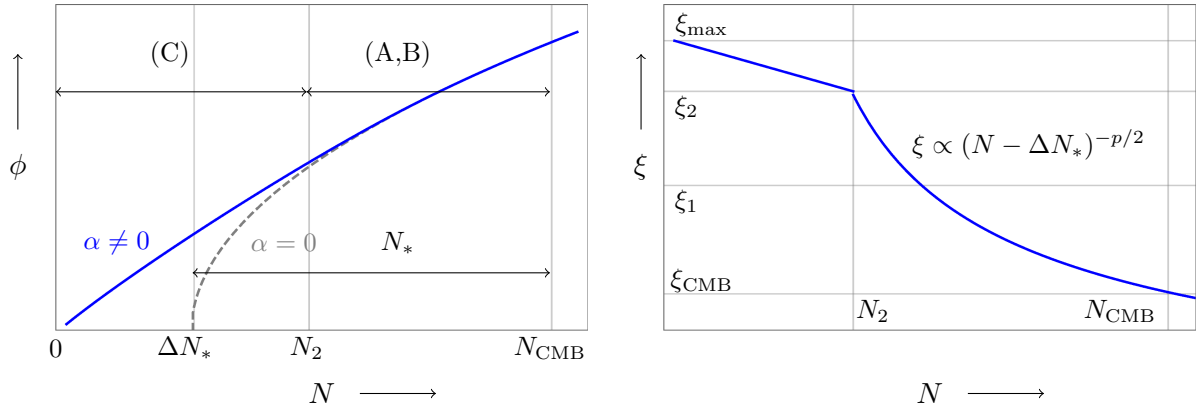

\noindent
 In this section, we will analytically examine the evolution of $\xi$ and the resulting universal features in the scalar and tensor power spectrum. The starting point is the equation of motion~\eqref{eq:eq_motion}, which, employing $dN = - H \, dt$, can be expressed as
 \begin{equation}
\label{eq:approx_eq_motion_2}
- \phi_{, N} + \frac{V_{,\phi}}{V}  = {\cal N} \, \frac{2.4}{9} \cdot 10^{-4} \left( \frac{\alpha}{\Lambda} \right) \frac{V}{\xi^4} e^{2 \pi \xi} \,.
\end{equation}
 We will distinguish three regimes: In regime (A), vacuum fluctuations dominate and the effects of the gauge fields are negligible. In regime (B), the contributions to the power spectra from the gauge fields overcome those stemming from the vacuum fluctuations. For both (A) and (B) we may neglect the gauge contribution on the righthand side of Eq.~\eqref{eq:approx_eq_motion_2}. In regime (C), backreaction effects of the gauge fields on the inflation dynamics become important, and we may instead neglect the first term on the lefthand side, corresponding to the Hubble friction. To facilitate orientation, a schematic overview of the evolution of $\phi$ and $\xi$ is depicted in Fig.~\ref{fig:phi_xi_schematic}. The meaning of marked values of $N$ and $\xi$ will be explained below.  For simplicity we set ${\cal N} = 1$ in the remainder of this section.

\subsection{Classifying the strength of the gauge fields. \label{sec:analytic1}}

\noindent\textbf{Vacuum dominated regime (A)}

\noindent The equation of motion reads:
\begin{equation}
- \phi_{, N} + \frac{V_{,\phi}}{V} \simeq 0,
\label{eq:motion0}
\end{equation}
and the amplitudes of the scalar and tensor power spectra are given by their vacuum contributions:
\begin{equation}
\Delta_s^2 = \frac{H^2}{8 \pi^2 \epsilon_V} \bigg|_{N_* = N_{CMB} - \Delta N_*} \,, \quad \Omega_{GW} = \frac{\Omega_{R,0} H^2}{12 \pi^2}    
= \frac{4}{3} \Delta_s^2 \epsilon_V \Omega_{R,0} \bigg|_{N_* = N_{CMB} - \Delta N_*} \,,
\label{eq:vacuumamplitudes}
\end{equation}
with $N_\text{CMB} \simeq 50 - 60$ denoting the number of e-folds of inflation elapsed since the CMB scales exited the horizon, $N_*$ denoting the amount of these required to cover the same distance in field space for $\alpha = 0$ and consequently $\Delta N_*$ denoting the number of additional e-folds of inflation due to the friction of the gauge fields, cf.\ Fig.~\ref{fig:phi_xi_schematic}. We will define the end of this regime based on the GW spectrum \eqref{eq:OmegaGW}, i.e.\
\begin{equation}
\xi < \xi_1 \quad \text{with} \quad \frac{e^{4 \pi \xi_1}}{\xi_1^6} = \left(4.3 \cdot 10^{-7} H_1^2 \right)^{-1} \,.
\label{eq:analytical_xi1}
\end{equation}
where we can estimate $H_1$ as the Hubble parameter at the CMB scale,
\begin{equation}
H_1^2 \simeq \frac{\pi^2}{2} \Delta_s^2 \frac{16 \beta_p}{(N_\text{CMB})^p} \,.
\end{equation}
For a given value of $\xi_\text{CMB}$ at CMB scales, this value of $\xi$ can be translated into a value for $N$,
\begin{equation}
\frac{\xi_1}{\xi_\text{CMB}} = \left( \frac{N_\text{CMB} - \Delta N_*}{N_1 - \Delta N_*} \right)^{p/2} \,.
\label{eq:xigrowth2}
\end{equation}
which finally can be translated into a frequency according to Eq.~\eqref{eq:Nf}.

\vspace{1cm}

\noindent\textbf{power law regime (B)}

\noindent As $\xi > \xi_1$, the contribution from the gauge fields begins to dominate the GW spectrum. In this regime, as long as the additional friction in Eq.~\eqref{eq:eq_motion} is small, the  equation of motion Eq.~\eqref{eq:motion0} still holds and the evolution of $\xi$ is still well described by Eq.~\eqref{eq:xigrowth2}, leading to a strongly blue scalar and tensor spectrum. 
Regime (B) lasts until the gauge field friction term  in Eq.~\eqref{eq:approx_eq_motion_2} overcomes the Hubble friction, i.e.\ for
\begin{equation}
\xi_1 < \xi < \xi_2, \quad 
\frac{ e^{2 \pi \xi_2}}{\xi_2^{5} }\simeq \left( \frac{\alpha}{\Lambda} \right)^{-2} \left[ 0.4 \cdot 10^{-4} H^2 \right]^{-1}\,.
\label{eq:knee_equation}
\end{equation}
This value for $\xi$ can be translated into a value for $N$ and $f$ according to Eqs.~\eqref{eq:xigrowth2} and \eqref{eq:Nf}, after evaluating the (model dependent) value of $H$ at this point.\\

\noindent
The corresponding value for $\phi_2 = \phi(N_2)$ (and analogously $\phi_1$) can be obtained by solving the slow-roll equation of motion Eq.~\eqref{eq:motion0}.
The constant of integration, i.e.\ the value of $\phi(N_*)$ can be calculated from the $\alpha = 0$ dynamics taking into account the shift $\Delta N_*$.

\vspace{1cm}

\noindent\textbf{strong gauge field regime (C)}

\noindent Finally we consider the regime $\xi_2 < \xi$. The non-linear friction term in Eq.~\eqref{eq:approx_eq_motion_2} becomes dominant, leading to $|\phi_{,N}| \ll |V_{, \phi}/V|$,
\begin{equation}
 \frac{V_{,\phi}}{V}  \simeq \frac{0.8}{3} \cdot 10^{-4} \left( \frac{\alpha}{\Lambda} \right) \frac{V}{\xi^4} e^{2 \pi \xi} \,.
 \label{eq:eomC}
\end{equation}
During slow-roll inflation, the lefthand side of this equation changes only moderately with time, bounded from above by the vacuum solution $V'/V \sim 1/N^{p/2}$. This implies that $\xi$ can grow at most logarithmically with decreasing $N$ in this regime.\footnote{A useful analogy is an object falling in some medium. As in this case, the friction-dependent velocity term leads to an approximately constant asymptotic velocity $\phi_{,N} \sim \xi$. Note that the friction term in this regime is stronger than required by the usual assumptions made in slow-roll inflation, which lead to neglecting the $\ddot \phi$-term.} This lasts until inflation ends, i.e.\ as long as $0 < \ddot a $, that is $\epsilon_H = |\dot H|/H^2 < 1$. Saturating this bound yields an upper bound for $\xi$: With $\dot H \simeq V' \dot \phi/(6 H)$ and hence $V' \lesssim 3 H^2 \alpha/(\xi \Lambda)$, the equation of motion~\eqref{eq:eomC} implies
\begin{equation}
\xi < \xi_\text{max} \,, \quad \frac{e^{2 \pi \xi_\text{max}}}{\xi_\text{max}^3} \lesssim \frac{3}{{\cal N} \cdot 2.4 \cdot 10^{-4} H^2} \,.
\label{eq:xi_end}
\end{equation}
In the parameter regime of interest, $\xi_\text{max} > 1$, this implies that low-scale models of inflation allow for larger values of $\xi$ and hence for stronger effects due to the presence of gauge fields.\\

\noindent
The bound \eqref{eq:xi_end} has interesting consequences for the scalar and tensor power spectrum. In this regime, the scalar power spectrum is given by Eq.~\eqref{eq:scalar2}, i.e.\ it is proportional to $1/\xi^2$. From Eq.~\eqref{eq:xi_end} we hence see that the scalar power at small scales (a potentially dangerous source of primordial black holes) is suppressed in low-scale models of inflation. Note in particular that the bound~\eqref{eq:xi_end} and hence the amplitude of the scalar power spectrum in this regime are independent of the parameter $\alpha$.\\

\noindent
Moreover, inserting Eq.~\eqref{eq:xi_end} into Eq.~\eqref{eq:OmegaGW} implies an absolute upper bound on $\Omega_{GW}$,
\begin{equation}
\Omega_{GW} h^2 \lesssim 2.4 \cdot 10^{-5} {\cal N}^{-1} \,,
\label{eq:OmegaMax}
\end{equation}
which holds independently of the inflation model. Here, as in Eq.~\eqref{eq:xi_end}, we have exceptionally re-introduced the parameter ${\cal N}$ to emphasize the parameter dependence of this bound. Since this bound is saturated at the end of inflation, this maximal value moreover is reached at a universal frequency. Inserting $N = 0$ into Eq.~\eqref{eq:Nf} yields
\begin{equation}
f_\text{max} \simeq 3.6 \cdot 10^8~\text{Hz} \,.
\end{equation}

\noindent
Next let us consider the inflationary dynamics in the strong gauge field regime. With $\xi$ approximately constant, we can estimate
\begin{equation}
\label{eq:phi_last}
\phi \simeq \bar{\phi}_{,N} N + \phi_0,
\end{equation}
with $|\bar \phi_{,N}| =  2 \bar \xi \Lambda/\alpha $, $\bar \xi = (\xi_\text{max} + \xi_2)/2 $. With this we can estimate the amount of e-folds in this strong gauge field regime,
\begin{equation}
\label{eq:N_estimate}
N_2 = (\phi_2 - \phi_0) \frac{\alpha}{2 \Lambda \bar{\xi}},
\end{equation}
with $\phi_0$ denoting the value of the inflaton field at the end of inflation, to good approximation determined by $\epsilon_V = 1$. As above, the value of $N$ can be translated into the corresponding frequency. Moreover, by solving the vacuum slow-roll dynamics between $\phi_2$ and $\phi_0$ using Eq.~\eqref{eq:motion0}, i.e.\ setting $\alpha = 0$, we can finally determine $\Delta N_*$ as
\begin{equation}
\Delta N_* = N_2 - N_2^0\,,
\label{eq:N2}
\end{equation}
with $N_2^0$ the numer of e-folds elapsed between $\phi_2$ and $\phi_0$ for $\alpha = 0$. Reinserting this value into the above expressions of regime (A) and (B), we obtain an analytical description of all the relevant points in scalar and tensor spectrum.

\subsection{The scalar and tensor spectra}
\label{sec:scalar_tensor}

The scalar power spectrum is observed to be nearly scale-invariant around the CMB pivot scale with an amplitude of $\Delta_s^2 \simeq 2.2 \cdot 10^{-9}$. Within the framework of Eq.~\eqref{eq:Nparameterization}, the tilt of this spectrum is obtained as
\begin{equation}
n_s \simeq 1 - \frac{{\cal O}(1)}{N_*}\,,
\label{eq:nsN}
\end{equation}
with the ${\cal O}(1)$ - factor depending on the choice of inflation model. Including the effects of the gauge field production, $N_* = N_\text{CMB} - \Delta N_* < N_\text{CMB}$, and hence the spectral index decreases compared to the $\alpha = 0$ case. The observed value of $n_s$ thus imposes an upper bound on $\Delta N_*$. The precise value depends on the ${\cal O}(1)$ in Eq.~\eqref{eq:nsN}, but typically we find $\Delta N_* \lesssim 10 - 20$. Consequently, this constrains the value of $N_2$ through Eq.~\eqref{eq:N2} and implies an upper bound on $\alpha/\Lambda$ in Eq.~\eqref{eq:N_estimate}. \\

\begin{figure}
\centering
\begin{tikzpicture}
\node at (0,0) { \includegraphics[width=0.65 \textwidth]{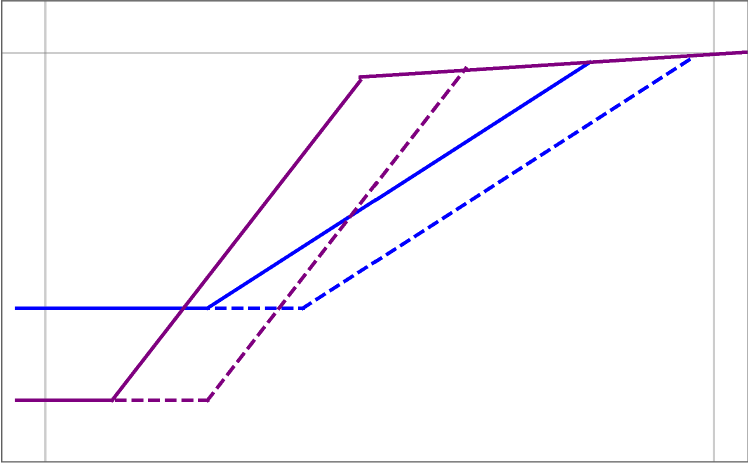}};
\node at (0, -3.7) {log $f$};
\node[rotate = 90] at (-5.7,-0.3) {log $\Omega_\text{GW}$};
\draw[->] (-5.7, 1) -- (-5.7,2);
\draw[->] (1, -3.7) -- (2, -3.7);
\node at (-3.5,1) {\textcolor{blue}{$p_1$}\,$<$\,\textcolor{purple}{$p_2$}};
\node at (-4,-2.8) {\small $f_\text{CMB}$};
\node at (4, -2.8) {\small $f_\text{max}$};
\node at (-3.5, 2.7) {\small $\Omega_\text{max}$};
\node at (-2, -2.55) {\small \textcolor{purple}{$(f_1, \Omega_\text{GW,1})$}};
\node at (0.2, -1.2) {\small \textcolor{blue}{$(f_1, \Omega_\text{GW,1})$}};
\node at (-1.5, 2.0) {\small \textcolor{purple}{$(f_2, \Omega_\text{GW,2})$}};
\node at (3.5, 2.7)  {\small \textcolor{blue}{$(f_2, \Omega_\text{GW,2})$}};
\draw[<-, purple] (-0.9,1)--(0.2,1);
\node at (-0.2, 1.2) {\small \textcolor{purple}{$\alpha$}};
\draw[<-, blue] (1,1)--(2,1);
\node at (1.7, 1.2) {\small \textcolor{blue}{$\alpha$}};
\end{tikzpicture}
\caption{Schematic view of the gravitational wave spectrum for two different values of $p$ in Eq.~\eqref{eq:Nparameterization} and two different values of the inflaton - gauge field coupling $\alpha/\Lambda$.}
\label{fig:Omega_schematic}
\end{figure}
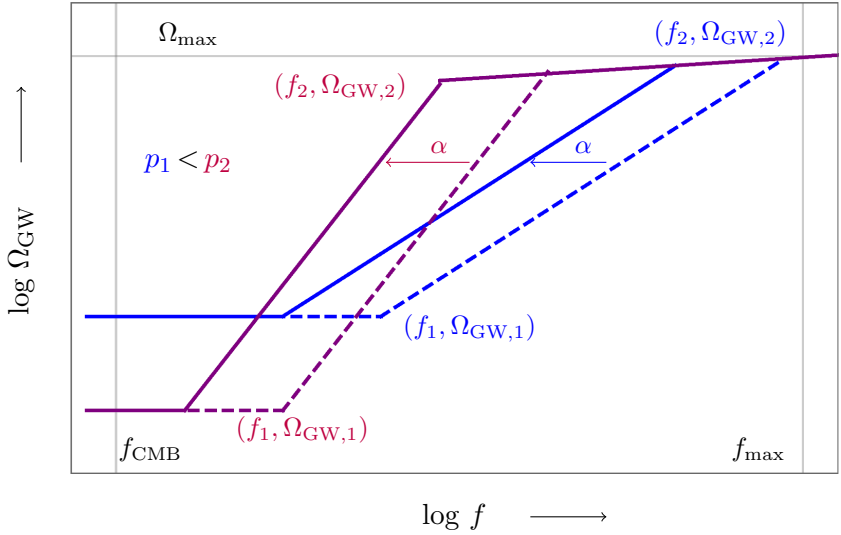

\noindent
Equipped with the results of Sec.~\ref{sec:analytic1}, we now turn to the prediction of the GW spectrum $\Omega_\text{GW}$, see also the schematic depiction in Fig.~\ref{fig:Omega_schematic}. Here we show the GW spectrum for two different values of $p$ with $p_1$ (blue) smaller than $p_2$ (purple). The dashed curves refer to reducing the value of $\alpha/\Lambda$ with respect to the corresponding solid curves.\\

\noindent
At very small frequencies, corresponding to the CMB scales, the amplitude is governed by the first slow-roll parameter and is hence to good approximation proportional to $\beta_p/N_\text{CMB}^p$. cf.\ Eq.~\eqref{eq:vacuumamplitudes} (regime A). At very large frequencies, corresponding to scales exiting the horizon towards the end of inflation, the universal value $\Omega_\text{max}$ is slowly approached, cf.\ Eq.~\eqref{eq:OmegaMax} (regime C). In between, there is a steep increase of the spectrum, governed by the $1/N^{p/2}$ growth of $\xi$ (regime B).
This increase last from $\xi = \xi_1$ to $\xi = \xi_2$, cf.\ Eqs.~\eqref{eq:analytical_xi1} and \eqref{eq:knee_equation}. Inflation models with a higher value of $p$ in Eq.~\eqref{eq:Nparameterization} correspond to a lower scale of inflation $H_1$ thus leading to a smaller value of $\xi_1$.\footnote{In the parameter regime we are interested in, $e^{2 \pi \xi}/\xi^3$ is monotonically incrasing with $\xi$.} For a fixed value of $\xi_\text{CMB}$, this shifts $f_1$ to smaller values, as illustrated in Fig.~\eqref{fig:Omega_schematic}.  Additionally, a larger value of $p$ implies a steeper slope of the spectrum between $f_1$ and $f_2$, due to the faster growth of $\xi$. As a result, the plateau in the spectrum, corresponding to an approximately constant value of $\xi$, extends to lower frequencies in these models. This leads to the interesting conclusion that models with a lower (vacuum) tensor-to-scalar ratio $r = 16 \epsilon$ actually have the larger GW signal in this setup.
Besides the value of $p$, the second important parameter is $\xi_\text{CMB}$ or equivalently $\alpha/\Lambda$. As can be seen from Eq.~\eqref{eq:xigrowth2}, reducing this values corresponds to shifting $N_1$ (and correspondingly $N_2$) to smaller values, i.e.\ shifting $f_1$ and $f_2$ to higher frequencies. \\

\noindent
Finally, the moderate increase in $\Omega_\text{GW}$ between $f_2$ and $f_\text{max}$ may be traced back to the moderate incrase of $\xi$ in that region, compensated to some extent by a decreasing Hubble parameter. Comparing Eqs.~\eqref{eq:knee_equation} and \eqref{eq:xi_end} yields
\begin{equation}
 \frac{e^{2 \pi \xi_2}}{\xi_2^3} = \frac{\phi_{,N}^2(N_2)}{2} \frac{e^{2 \pi \xi_\text{max}}}{\xi_\text{max}^3} \,.
\end{equation}
With $\phi_{,N}^2 \sim (V'/V)^2 \sim \epsilon \sim r$, we note that low-scale models which allow for an early and steep rise of the spectrum at the same time require a smaller value of $\xi_2$. \\

\noindent
In summary, $p$ controls the slope of the strong increase in the scalar and tensor spectrum, $\alpha/\Lambda$ shifts the entire spectrum horizontally and $\beta_p$ vertically shifts the vacuum part of the spectrum. It is the interplay of these three parameters which controls the detectability of the GW signal.

%% file: models.tex
\label{sec:explicit_models}
In this section we show the results obtained when we consider some specific inflationary potentials. Following a phenomenological approach, we study different types of inflationary potentials. Here, we are only interested in the behaviour of these potentials during the final 60 e-folds of inflation, constructing a possible UV-completion taking care of an appropriate shape of the potential beyond this region is beyond the scope of this paper. In particular we are interested in observing the differences in the GW production as we consider different inflationary models. As we have discussed in the previous sections, we expect the shape of the GW spectrum to be particularly sensitive to $\alpha/\Lambda$ and $p$. As most of the models discussed in literature \cite{Barnaby:2010vf,Barnaby:2011vw,Linde:2012bt} have $p = 1$, in this work we are interested in considering some models with a different value for this parameter. To produce a classification of models in terms of $p$ we use the parametrization of Eq.~\eqref{eq:Nparameterization}. It is then useful to recall the approximate relationship between the potential $V(\phi)$ and the number of e-folds $N$ :
\begin{equation}
\label{eq:models_N}
\frac{\textrm{d} N}{\textrm{d} \phi} \simeq \left( \frac{\textrm{d} \ln V(\phi)}{\textrm{d} \phi}  \right)^{-1} .
\end{equation}
By differentiating Eq.~\eqref{eq:Nparameterization} and substituting into Eq.~\eqref{eq:models_N} we can obtain the differential equation: 
\begin{equation}
\label{eq:models_differential}
\epsilon_{V,\phi} = - \frac{p}{\sqrt{2} \beta_p^{ \ \frac{1}{p}}} \epsilon_V^{\ \frac{p+2}{2p}}.
\end{equation}
To solve this differential equation we must distinguish $p = 2$ from all the other cases.
\begin{itemize}
	\item $p = 2$.
	In this case Eq.~\eqref{eq:models_differential} reduces to:
	\begin{equation}
	\epsilon_{V,\phi} = - \frac{2}{\sqrt{2 \beta_p } } \epsilon_V,
	\end{equation}
	whose solution is given by:
	\begin{equation}
	\label{eq:models_epsilon_starobinsky}
	\epsilon_V \simeq \exp\left( -\sqrt{\frac{2}{\beta_p}} \phi \right).
	\end{equation}
	\item $p \neq 2$.
	In this case the solution of the differential equation is given by:
	\begin{equation}
	\label{eq:models_epsilon_general}
	\epsilon_V \simeq \left( -\frac{(p-2)}{\sqrt{8}\beta_p^{\ \frac{1}{p}}} \phi \right)^{\frac{2p}{p-2}}.
	\end{equation}
\end{itemize}
To complete our classification we can use the expression of $\epsilon_V$ given in Eq.~\eqref{eq:epsilon_HV} to constrain the expression for $V(\phi)$. In particular it is easy to show that for chaotic models \cite{Linde:1983gd} with potential:
\begin{equation}
\label{eq:models_chaotic_potentials}
V (\phi)= V_0 \ \phi^q ,
\end{equation}
we get $p=1$. To consider different values of $p$ we can start by noticing that Starobinsky model \cite{Starobinsky:1980te}:
\begin{equation}
\label{eq:models_starobinsky_potential}
V(\phi) =  V_0 \left(1 - e^{- \sqrt{\frac{2}{3}} \phi}\right)^2 ,
\end{equation}
corresponds to $p =2$. It is interesting to point out that this result holds for a more general class of models\footnote{In the rest of this paper we always use the term Starobinsky model to refer to this general class of models.} with:
\begin{equation}
\label{eq:models_starobinsky_potentials}
V(\phi) = V_0 \left(1 - e^{- \gamma \phi}\right)^2 .
\end{equation}
Finally we notice that Hilltop models \cite{Boubekeur:2005zm} with a potential given by:
\begin{equation}
\label{eq:models_hilltop_potentials}
V(\phi) =  V_0 \left[1 - \left (\frac{\phi}{v} \right)^q \right]^2,
\end{equation}
correspond to $p = 2 (q-1)/(q-2)$. \\

\subsection{CMB constraints and numerical results.}
\label{sec:Constraints_and_results}
Before turning to the particular models with different values for $p$, it is useful to recall the conditions that must be satisfied in all of these cases. In particular all the models must be in agreement with three constraints coming from CMB observations. It is crucial to stress that all the quantities appearing in these constraints should be evaluated at horizon crossing. According to the discussion of Sec.~\ref{sec:analytical}, this corresponds to using the complete evolution~\eqref{eq:approx_eq_motion_2} and evaluating quantities at $N = N_{CMB}$. However it is useful to point out that if $\Delta N_* $ is considerably smaller than $ N_\text{CMB} \simeq 60$, in this regime the gauge field contribution is fairly negligible. Under this assumption, a good estimate of the following constraints can be obtained by using the standard vacuum slow-roll evolution~\eqref{eq:motion0} and evaluating quantities at $N = N_*$. As in our numerical simulations we have used the complete evolution, we provide these constraints in the most general form. However, for the estimates of Sec.~\ref{sec:starobinsky}, we use their approximated expression in terms of $N_*$.

\begin{itemize}
\item \textbf{COBE Normalization}: It sets the value of the scalar power spectrum at the CMB scales. This condition can be used to fix a constraint on the inflationary potential. In particular we have \cite{Ade:2015lrj}: 
	\begin{equation}
	\label{eq:models_COBE}
		 \left. \Delta_s^2\right|_{N = N_{CMB}} = (2.21 \pm 0.07) \cdot 10^{-9} \,.
	\end{equation}
	\item \textbf{Planck measurements}: These further constrain the spectral index $n_s$, the running of the spectral index $\alpha_s$ and the tensor-to-scalar ratio $r$, defined as
	\begin{equation}
n_s - 1 = \frac{d \ln \Delta_s^2}{d\ln k} \,, \qquad \alpha_s = \frac{d n_s}{d \ln k} \,, \qquad r = \frac{\Delta_t^2}{\Delta_s^2} \,.
	\end{equation}
	The constraints on these parameters from the Planck mission \cite{Ade:2015lrj} read (at $68\%$ CL for $n_s$ and $\alpha_s$, $95\%$ CL for $r$):
	\begin{equation}
		\label{eq:models_PLANCK}
n_s = 0.9645 \pm 0.0049\,, \qquad \alpha_s = -0.0057 \pm 0.0071\,, \qquad r < 0.10 \,.
	\end{equation}
	In slow-roll approximation and for a negligible gauge field contribution at the CMB scales, the quantities above are given by:
	\begin{equation}
		\label{eq:models_PLANCK2}
		 \Delta_s^2 = \frac{1}{24 \pi^2} \frac{V(\phi)}{\epsilon_V(\phi)}\,, \qquad n_s \simeq 1  +2\eta_V - 6 \epsilon_V   , \qquad  r \simeq 16 \epsilon_V\,,
	\end{equation}
	where $\epsilon_V $ is defined in Eq.~\eqref{eq:epsilon_HV} and $\eta_V$ is defined as $\eta_V = V_{,\phi \phi}/V$. It is useful to express $\eta_V$ as:
	\begin{equation}
	\label{eq:eta_N}
	\eta_V = \frac{1}{2} \frac{\textrm{d} \ln \epsilon_V }{\textrm{d} N} + 2\epsilon_V,
	\end{equation} 
	yielding~\cite{Mukhanov:2013tua}:
	\begin{equation}
	n_s = 1 - \frac{p}{N} - 6 \epsilon \,.
	\end{equation}
	For $p>1$, the term proportional to $\epsilon$ is negligible, indicating that $n_s \sim 0.96$ suggests $p < 2.4$ for $N_\text{*} < 60$. 

	\item \textbf{Small non gaussianities}: As discussed in \cite{Anber:2012du,Barnaby:2010vf,Barnaby:2011vw,Barnaby:2011qe,Linde:2012bt}, to respect the constraints on small primordial non gaussianities we need $\xi_{CMB} \equiv \left. \xi\right|_{N=N_{CMB}} \lesssim 2.5$. This implies:
	\begin{equation}
	\label{eq:models_NG}
		\xi_{CMB} = \frac{\alpha}{2 \Lambda} \left|\frac{\dot{\phi}}{H}\right|_{N=N_{CMB}} \lesssim 2.5.
	\end{equation}
	More details on the derivation of this constraint are given in Sec.~\ref{sec:non_gaussianities}.
\end{itemize}

\begin{figure}[h]
\centering
{\includegraphics[width=.55\columnwidth]{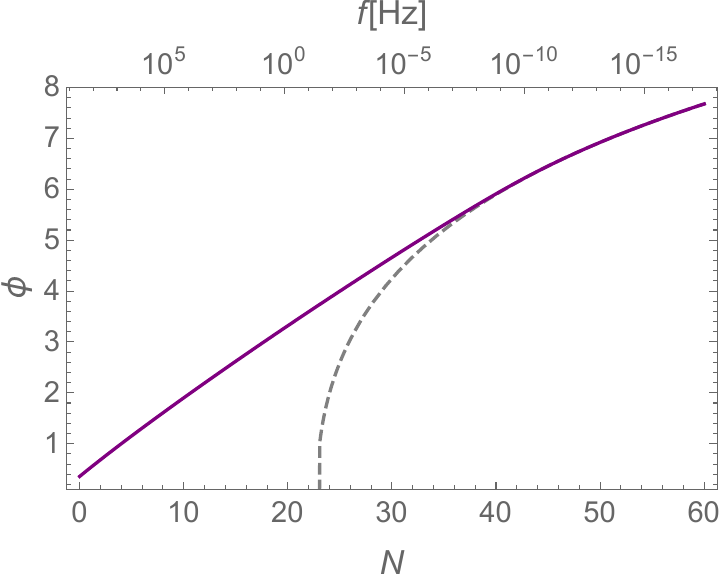}}
\caption{Evolution of inflaton field $\phi$ as a function of $N$ and $f$ (see Eq~.\eqref{eq:Nf}) for the Starobinsky model with (solid line) and without (dashed line) the non-minimal interaction with the gauge fields. We show the evolution for the case with $\alpha /\Lambda \sim 75 $, $\gamma~=~0.3$, $V_0 \simeq 1.525 \cdot 10^{-9}$.} 
\label{fig:phi_comparison_staro}
\end{figure}

\noindent
The evolution of $\phi$ for the Starobinsky model is shown in Fig.~\ref{fig:phi_comparison_staro}. These and the following results have been obtained by numerically solving the slow-roll equation of motion~\eqref{eq:approx_eq_motion_2} for a fixed parameter point.\footnote{Here we have set $N_\text{CMB} = 60$. As discussed in Sec.~\ref{sec:discussion}, there is a degeneracy between the choice of $N_\text{CMB}$ and $\alpha/\Lambda$.} As expected, the coupling between the gauge field and the inflaton only effects the last part of the evolution. In particular it slows down the variation of $\phi$ and stretches the length of inflationary phase. Similar plots for models with different values for $p$ are presented in Appendix~\ref{sec:Appendix}.\\

\noindent
In Figs.~\ref{fig:xi_all},  \ref{fig:DeltaS_all} and \ref{fig:GW_all} we present $\xi$, $\Delta_s^2$ and $\Omega_{GW}$ respectively for the models with $p = 1,2,3,4$.  As anticipated in Sec.~\ref{sec:analytical}, the plots for $\xi$ for all the different models are approximately resembling the plot of Fig.~\ref{fig:phi_xi_schematic}. In particular, we can immediately notice that the plot of Fig.~\ref{fig:xi_all} is in agreement with the estimate of Eq.~\eqref{eq:xi_end}: namely, models with lower values for $H$ (i.e. Hilltop models with $p=3,4$) have a bigger value for $\xi_{\text{max}}$. We can also notice that the value of $\xi_\text{CMB}$ for the different models are respecting the condition of Eq.~\eqref{eq:models_NG}. \\

\begin{figure}[h]
\centering
{\includegraphics[width=.6\columnwidth]{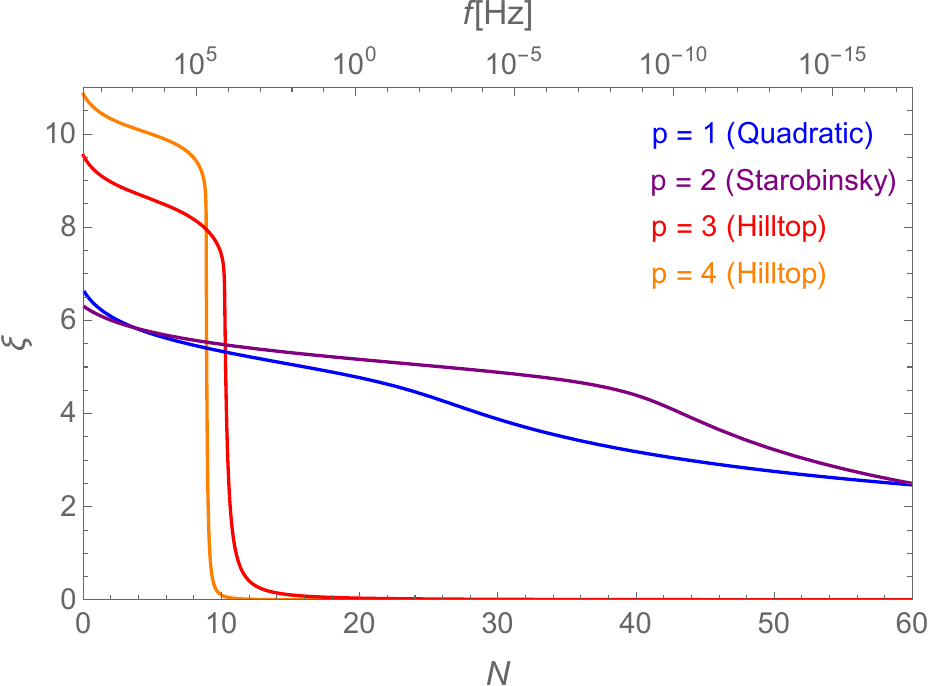}}
\caption{Evolution of the parameter $\xi$ governing the strength of the gauge interactions for models with different values of $p$ as defined in Eq.~\eqref{eq:Nparameterization}. The parameters for the Starobinsky model are as in Fig.~\ref{fig:phi_comparison_staro}, the parameters for the other models are listed in App.~\ref{sec:Appendix}.}
\label{fig:xi_all}
\end{figure}

\noindent
As it is possible to see from Fig.~\ref{fig:DeltaS_all}, we fix the parameters of our models in order to fit the COBE normalization at $N \simeq 60$. Moreover, all of these spectra are nearly flat in order to respect the constraints of Eq.~\eqref{eq:models_PLANCK}. In agreement with the estimate of \cite{Linde:2012bt}, the value of $\Delta_s^2$ on small scales is proportional to $\xi^{-2}$ and thus the corresponding value for the Hilltop models tends to be smaller. It is worth to point out that all of the models considered in this paper are in tension with the estimated PBH bound of \cite{Linde:2012bt} when we restrict to the case ${\cal N} = 1$.  As this discrepancy is however only by a $\mathcal{ O}(1)$ factor, it can both be addressed by taking into account the theoretical uncertainties in the PBH bound (see also Sec.~\ref{sec:bounds}) or by considering models with $ {\cal N}>1$, see Fig.~\ref{fig:several_gauge_plots}. As evident from the figure, the scalar spectrum for the Hilltop models i.e. $p =3,4$ presents a much steeper decrease in the first part of the evolution with respect the other models, as predicted by Eq.~\eqref{eq:vacuumamplitudes}, $\epsilon_V \simeq N^{-p}$. \\

\begin{table}[hb]
\centering
\begin{tabular}{llccc}
name & full name & number of arms & armlength [Gm] & lifetime [yr] \\ \hline
C1 & L6A5M5N2 & 3 & 5 & 5 \\
C2 & L6A1M5N2& 3 & 1 & 5 \\
C3 & L4A2M5N2 & 2 & 2 & 5 \\
C4 & L4A1M2N1& 2 & 1 & 2 
\end{tabular}
\caption{Configurations of the planned space-based GW mission eLISA considered in this paper.}
\label{tab:lisa}
\end{table}

\noindent
The GW spectrum for all the models considered in this paper is shown in Fig.~\ref{fig:GW_all}. In agreement with the discussion of Sec.~\ref{sec:scalar_tensor}, all of these models are reproducing the schematic behavior shown in Fig.~\ref{fig:Omega_schematic}. In particular we can always appreciate two abrupt changes in the slope of the curves for two different values of the frequency. Further we depict in Fig.~\ref{fig:GW_all} the sensitivity curves of a selection of current (solid lines) and upcoming (dashed lines) direct GW detectors. Representing the millisecond pulsar timing arrays covering frequencies around $10^{-10}$~Hz, we show the constraint depicted in Ref.~\cite{Smith:2005mm}, the update from EPTA~\cite{vanHaasteren:2011ni} and the expected sensitivity of SKA~\cite{Kramer:2004rwa}. This is followed by space-based GW interferometers in the milli-Hz range (eLISA~\cite{Petiteau}) and ground-based detectors sensitive at a few 10 Hz (LIGO/VIRGO~\cite{TheLIGOScientific:2016wyq}). For eLISA, we depict the sensitivity curves for the four configurations listed in Tab.~\ref{tab:lisa}. For LIGO, we depict the current bound O1:2015-16, as well as the expected sensitivities for the runs O2:2016-17 an O5:2020-22. Note that the GW signal generated in this setup is maximally chiral, distinguishing it from other stochastic GW backgrounds.\\

\noindent
Fig.~\ref{fig:GW_all} clearly shows that, with a particular parameter choice, both the Quadratic and Starobinsky model can be in the observable window for eLISA and advanced LIGO. Moreover, this particular set of parameters for the Starobinsky model happens to produce a GW spectrum that can be observed by the milli-second pulsar timing. On the contrary, the GW spectrum for the Hilltop models is well outside the observable windows for all of these experiments.\\

\begin{figure}[p]
\centering
{\includegraphics[width=.8\columnwidth]{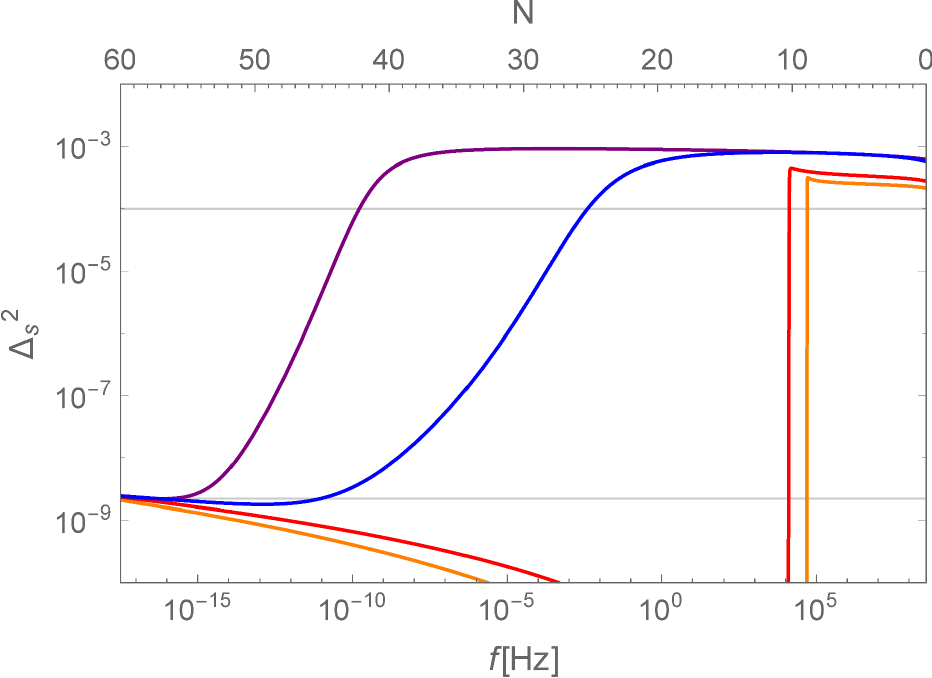}}
\caption{Power spectrum of scalar perturbations for all the models with the same parameters and color code of Fig.~\ref{fig:xi_all}. The upper horizontal line estimates the PBH bound, the lower one indicates the COBE normalization.}
\label{fig:DeltaS_all}
\end{figure}

\begin{figure}[p]
\centering
{\includegraphics[width=.8\columnwidth]{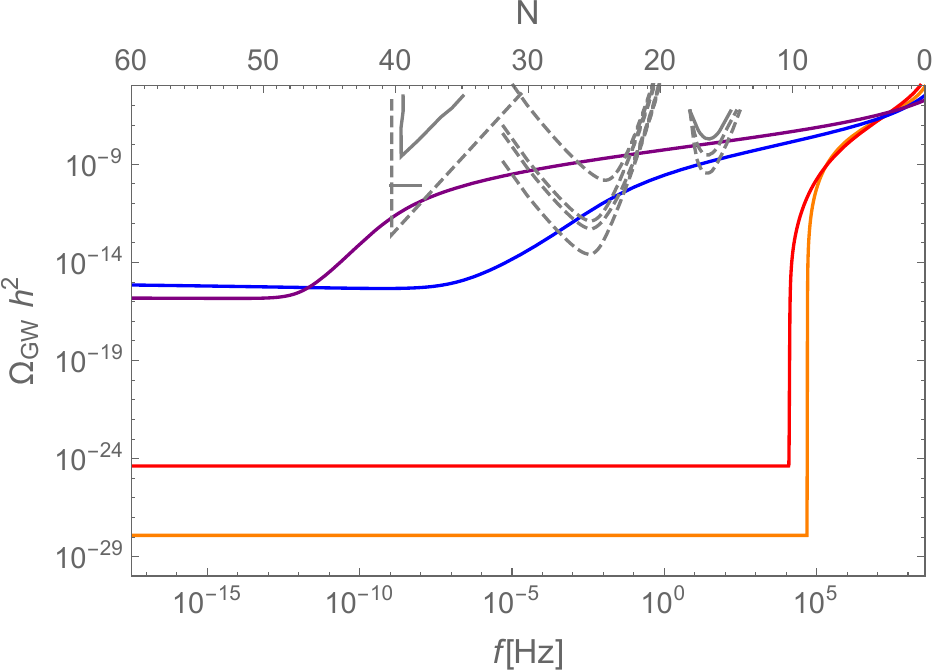}}
\caption{Gravitational wave spectrum for all the models with the same parameters and color code of Fig.~\ref{fig:xi_all}. We are also showing the sensitivity curves for (from left to right): milli-second pulsar timing, eLISA, advanced LIGO. Current bounds are denoted by solid lines, expected sensitivities of upcoming experiments by dashed lines. See main text for details.}
\label{fig:GW_all}
\end{figure}

\noindent
In the remainder of this section we discuss the particular parameter choice for the case of Starobinsky model with potential given by Eq.~\eqref{eq:models_starobinsky_potential}, as this seams to be the most interesting class of models from the point of view of potentially observable signatures. In particular, we want to show that the estimates presented in Sec.~\ref{sec:analytical} are coherent with the numerical results that we have obtained. An analogous treatment for all the other models shown in the plots of this section is presented in Appendix~\ref{sec:Appendix}. It is important to stress that through these estimates we have been able to fix the parameters in order to maximize the GW signal without violating the constraints of Eq.~\eqref{eq:models_PLANCK}, Eq.~\eqref{eq:models_COBE} and Eq.~\eqref{eq:models_NG}. It is then possible to reduce the signal basically by reducing the value of $\alpha/\Lambda$, thus shift the rise in the spectrum to larger frequencies. A numerical scan of the parameter space will be presented in Sec.~\ref{sec:discussion}, the execution of which is however greatly facilitated by the use the analytical expressions obtained below to determine the parameter space of interest.

\subsection{Comparison with analytical results.}
\label{sec:starobinsky}

To discuss the case $p =2$ we consider the generalized version of the Starobinky potential i.e.
\begin{equation}
\label{eq:starobinsky_potential}
V = \frac{3}{4} V_0 \left(1 -   e^{- \gamma \phi}\right)^2 .
\end{equation}
For this class of models the lowest order expression for the slow-roll parameters is given by:
\begin{equation}
\label{eq:models_starobinsky_slowroll}
\epsilon_V = \frac{1}{2 \gamma^2 N^2 } \qquad \qquad \eta_V \sim  - \frac{1}{N}
\end{equation}
To fix the parameters of the model we can use Eq.~\eqref{eq:models_COBE}, Eq.~\eqref{eq:models_NG} and Eq.~\eqref{eq:models_PLANCK}. We start by imposing the latter to get : 
\begin{equation}
\label{eq:Planck_Starobinsky}
N_* \simeq 50, \qquad \qquad  \frac{8}{250} \lesssim \gamma^2 . 
\end{equation}
We can then impose the condition on $\xi_*$ to get :
\begin{equation}
\label{eq:xi_CMB_Starobinsky}
 \frac{\alpha}{\Lambda} \lesssim  250 \gamma.
\end{equation}
Finally the COBE normalization implies:
\begin{equation}
\label{eq:COBE_Starobinsky}
V_0 \simeq \gamma^{-2} \cdot 1.04 \cdot 10^{-10} \ , \qquad \rightarrow \qquad  V_0 \lesssim \left (\frac{\alpha}{\Lambda} \right)^{-2} \cdot 6.5 \cdot 10^{-6} \ . 
\end{equation}
All of these conditions can be satisfied with an appropriate choice for the free parameter $\alpha /\Lambda$. In the following we show the results obtained by imposing $\alpha /\Lambda \sim 75 $, $\gamma = 0.3$, $V_0 = 1.525 \cdot 10^{-9}$. This choice for  $\alpha/\Lambda$ is motivated by the constraints of $n_s$ and $r$ set by Eq.~\eqref{eq:models_PLANCK}. In particular this corresponds to $n_s = 0.960, \  r = 0.036$ and $\alpha_s \equiv \textrm{d} n_s/\textrm{d} \ln k = 0.00352$. \\

\noindent
The value of $\xi_1$ can be estimated by substituting $V_0$ into Eq.~\eqref{eq:analytical_xi1}:
\begin{equation}
 	\label{eq:models_xi1_staro}
 	\xi_1^6 = 1.68 \cdot 10^{-16} \cdot \exp\left( 4 \pi \xi_1\right), \qquad \longrightarrow \qquad  \xi_1 \simeq 3.49.
 \end{equation} 
To estimate the value of $\xi_{\text{max}}$ we start by estimating the value of $V(\phi)$ at the end of inflation. For this purpose we simply assume $V(\phi_{\text{max}}) \simeq V_0$. We can then use Eq.~\eqref{eq:xi_end} to get:
\begin{equation}
\xi_{\text{max}}^{3} \simeq 2.77 \cdot 10^{-15} e^{2 \pi \xi_{\text{max}}}, \qquad \longrightarrow \qquad \xi_{\text{max}} \simeq 6.21.
\label{eq:models_ximax_staro}
\end{equation}
The value of $\xi_2$ can be estimated from Eq.~\eqref{eq:knee_equation}. Again we assume $V(\phi) \simeq V$ to get:
\begin{equation}
\xi_2^{5} \simeq 8.77 \cdot 10^{-11} e^{2 \pi \xi_2}, \qquad \longrightarrow \qquad \xi_2 \simeq 4.96, \qquad \phi_2 \simeq 6.52.
\label{eq:models_xi2_staro}
\end{equation}
To estimate $\phi_2$, we need an approximated expression for $V(\phi)$, for this purpose we can use Eq.~\eqref{eq:epsilon_HV} to get $\phi_2 \simeq 5.70$. Finally, as in first part of the evolution $\xi$ is given by Eq.~\eqref{eq:xigrowth}, we can estimate the value for $\xi_{*}$ by using Eq.~\eqref{eq:epsilon_HV}.\\

\noindent
The estimated values for $N_2$, and $N_1$ and $\Delta N_*$ are obtained by using Eq.~\eqref{eq:N_estimate}, Eq.~\eqref{eq:xigrowth2} and Eq.~\eqref{eq:N2}: $N_2 \simeq 38.27$, $\Delta N_* \simeq 17.01$, $N_1 \simeq 47.81$. The corresponding frequencies are given by Eq.~\eqref{eq:Nf}:
\begin{equation}
	f_1 \simeq 6.23 \cdot 10^{-13}~\text{ Hz}, \qquad \qquad f_2 \simeq 8.66\cdot 10^{-9}~\text{ Hz} .
\end{equation}
As it is possible to see from Fig.~\ref{fig:GW_all}, these estimates are approximatively matching the values for these two frequencies in the full numerical computation.

%% file: experimental.tex
In this section we discuss the constraints imposed by some experimental observations on the production of gauge field quanta during inflaton. The first part of this section is used to present a more detailed discussion on the non-gaussianity bound of Sec.~\ref{sec:explicit_models}. We proceed by reviewing the constraints imposed by the non-observation of Primordial Black Holes (PBHs) and by indirect searches for gravitational waves. We then considering the possibility of generating primordial magnetic fields at the end of inflation and conclude by commenting on the prospects of generating observable $\mu$-type distortions in the CMB.

\subsection*{Non gaussianities.}
\label{sec:non_gaussianities}
As the gauge field contribution to the scalar power spectrum is non-gaussian, it is strongly constrained at the CMB scales (and LSS scales). The dominant non-gaussian contribution is denoted with $f_{NL}^\text{equil}$. Plack measurements \cite{Ade:2015lrj} constrain this quantity to be $|f_{NL}^\text{equil}| < |-4 \pm 43|$ at 68$\%$ CL. As widely discussed in literature~\cite{Barnaby:2010vf,Barnaby:2011vw,Barnaby:2011qe,Linde:2012bt}, we can express the three point function as:
\begin{equation}
f_{NL}^\text{equil} \simeq 6.16 \cdot 10^{-16} e^{6 \pi \xi}/\xi^9 \,,
\label{eq:fNL}
\end{equation}
where we have exploited that $\Delta_s^2$ is governed by the vaccuum contribution and constrained by Eq.~\eqref{eq:models_COBE} at the horizon crossing of the CMB modes. This directly implies $\xi < 2.5$ (95 $\%$ CL, gaussian errors assumed). It is worth to stress that Eq.~\eqref{eq:fNL} only applies in the weak gauge regime. On the contary in the strong gauge field regime we have $f_{NL}^\text{equil} = - 1.3 \, \xi$~\cite{Anber:2012du}.

\subsection*{Primordial black holes.}
As discussed in the previous sections, the increase of $\xi$ towards the end of inflation implies a strong rise in the scalar power spectrum. These strong perturbations on small scales in turn can lead to the formation of primordial black holes. The non-observation of PBHs can be used to set some constrains on the fraction of energy going into PBHs at their formation, as a function of the PBH mass. Constraints over a wide mass range have been collected in \cite{Carr:2009jm}. We can roughly divide the PBHs into three categories:
\begin{itemize}
	\item PBHs with masses smaller than $10^{15}$~g. These PBHs have already evaporated and they can be detected observing their entropy production in the early universe.
	\item PBHs with masses around $10^{15}$~g. These PBHs would be evaporating today and thus they would leave signals in $\gamma$-rays.
	\item PBHs with masses bigger than $10^{15}$~g. As these PBHs would still be stable, they can be searched for in lensing experiments. 
\end{itemize}

\noindent
To estimate the the fraction of PBHs we follow the treatment of \cite{Linde:2012bt}. As usual we define $\zeta \equiv - H\delta\phi / \dot{\phi}$ and $\zeta_c$, critical value leading to black hole formation. Given $P(\zeta)$, probability distribution for $\zeta$, we can express $b$, fraction of space that can collapse and form a PBH, as:
\begin{equation}
b = \int_{\zeta_c}^\infty P(\zeta) d\zeta .
\end{equation}
Hawking evaporation and present day gravitational effects constrain $b$, leading to $b \lesssim 10^{-28} - 10^{-5} $ depending on the PBH mass range, with the strongest bounds coming from CMB anisotropies \cite{Carr:2009jm}. According to the discussion presented in \cite{Linde:2012bt}, this can be translated into a constraint on the scalar power spectrum $\Delta_s^2 \lesssim 1.3 \cdot 10^{-4} - 5.8 \cdot 10^{-3}$.\footnote{Note that this constraint is considerably stronger compared to the one obtained by assuming gaussian fluctuations, $\Delta_s^2 \lesssim 10^{-2}$.} Following the analysis of \cite{Linde:2012bt}, it is also possible to associate the typical mass of a PBH with a given scale of the scalar perturbations sourcing it (and hence a corresponding value of $N$) through:
\begin{equation}
M_{PBH} = \frac{4 \pi}{H} e^{a N},
\end{equation}
where $a = \{2,3\}$ is a coefficient depending on the efficiency of the reheating. Since the PBH bound is strong for relatively light PBHs, this puts a strong constraint on the amplitude of the scalar perturbations at the end of inflation. Using the approximation for $\Delta_s^2$ in the strong gauge field regime this constraint can be used to directly put a lower bound on $\xi_{\text{max}}$ i.e. $ \xi_{\text{max}} \gtrsim 14/\sqrt{\cal N}$.\\

\noindent
It should be stressed that the calculation of PBH formation in these models is based on the strong gauge field regime, and the approximations performed in calculating this bound are estimated to account for up to an order one factor in the final bound~\cite{Linde:2012bt}. Moreover, as we will return to in the following section, the large amplitude of the scalar perturbations reached in this regime indicates that higher orders in the perturbative expansion may not be completely negligible. Ignoring these also contributes to the theoretical error in this regime. For these reasons, we depict the value of the  bound derived in~\cite{Linde:2012bt} in Fig.~\ref{fig:DeltaS_all}, but consider all the models discussed in this paper (which all violate this bound by an order one factor) as still viable, see also the dedicated discussion on the theoretical uncertainties in Sec.~\ref{sec:discussion}.\\

\noindent Finally, note that the large scalar perturbations we obtain on small scales also source sizable second order tensor perturbations~\cite{Baumann:2007zm}. In the models considered in this paper, these are however subdominant compared to the leading order GW contribution calculated above.

\subsection*{CMB and BBN bounds on primordial GWs.}
A further bound on primordial gravitational waves is imposed by the CMB and BBN constraints on additional massless degrees of freedom. As GWs with frequencies larger then the corresponding horizon at CMB decoupling or BBN contribute to the radiation density of the Universe, constraints on their total energy density can be phrased in terms of the effective number of massless neutrino species $N_\text{eff}$ (SM value: $N_\text{eff} = 3.046$),
\begin{equation}
\int d (\ln f) \, \Omega_\text{GW} = \Omega_{R,0} \frac{7}{8} \left( \frac{4}{11}\right)^{4/3} (N_\text{eff} - 3.046) \,,
\end{equation}
where the integral is performed over all frequencies $ f \gtrsim 10^{-15}$~Hz ($f \gtrsim 10^{-10}$~Hz) for the CMB (BBN) bound, see eg.\ Refs.~\cite{Allen:1996vm,Meerburg:2015zua,Cabass:2015jwe}. Compared to current bounds from the CMB ($N_\text{eff} = 3.04 \pm 0.17$)~\cite{Ade:2015xua} and from BBN ($N_\text{eff} = 3.28 \pm 0.28$)~\cite{Cyburt:2015mya}, the Starobinsky-type inflation model depicted in Fig.~\ref{fig:GW_all} exceeds the $95 \%$ CL region by an ${\cal O}(1)$ factor (in fact, this statement holds for most of the parameter space of interest, see Fig.~\ref{fig:mudis}). 
This tension may be resolved by a better understanding of the theoretical uncertainties in the strong gauge field regime (see Sec.~\ref{sec:discussion}) or by increasing the number of $U(1)$ gauge groups in the theory, cf.\ Fig.~\ref{fig:several_gauge_plots}. It is also interesting to point out that a recent paper from Riess et al. \cite{Riess:2016jrr} argues for a higher effective number of massless neutrino species i.e. $\Delta N_\text{eff} \simeq 0.4 - 1$. If confirmed, this could help to resolve the present tension.

\subsection*{Primordial magnetic fields.}
As a pseudoscalar inflaton should couple to all the $U(1)$ gauge field in the theory, it can also couple to the SM electromagnetic one. The gauge field production discussed in this paper could hence generate primordial magnetic fields. The generation of primordial magnetic fields in the model of our interest has been widely discussed in literature, see e.g.\ \cite{Durrer:2010mq, Anber:2006xt,Caprini:2014mja, Fujita:2015iga,Green:2015fss}. The result of these analysis is that, except for particular configurations, the magnetic fields generated with this mechanisms turn out to be too weak to provide the seeds for the observed fields in galaxies and clusters.

 \subsection*{$\mu$-type distortions in the CMB}

\begin{figure}[t]
\centering
\includegraphics[width=.45\columnwidth]{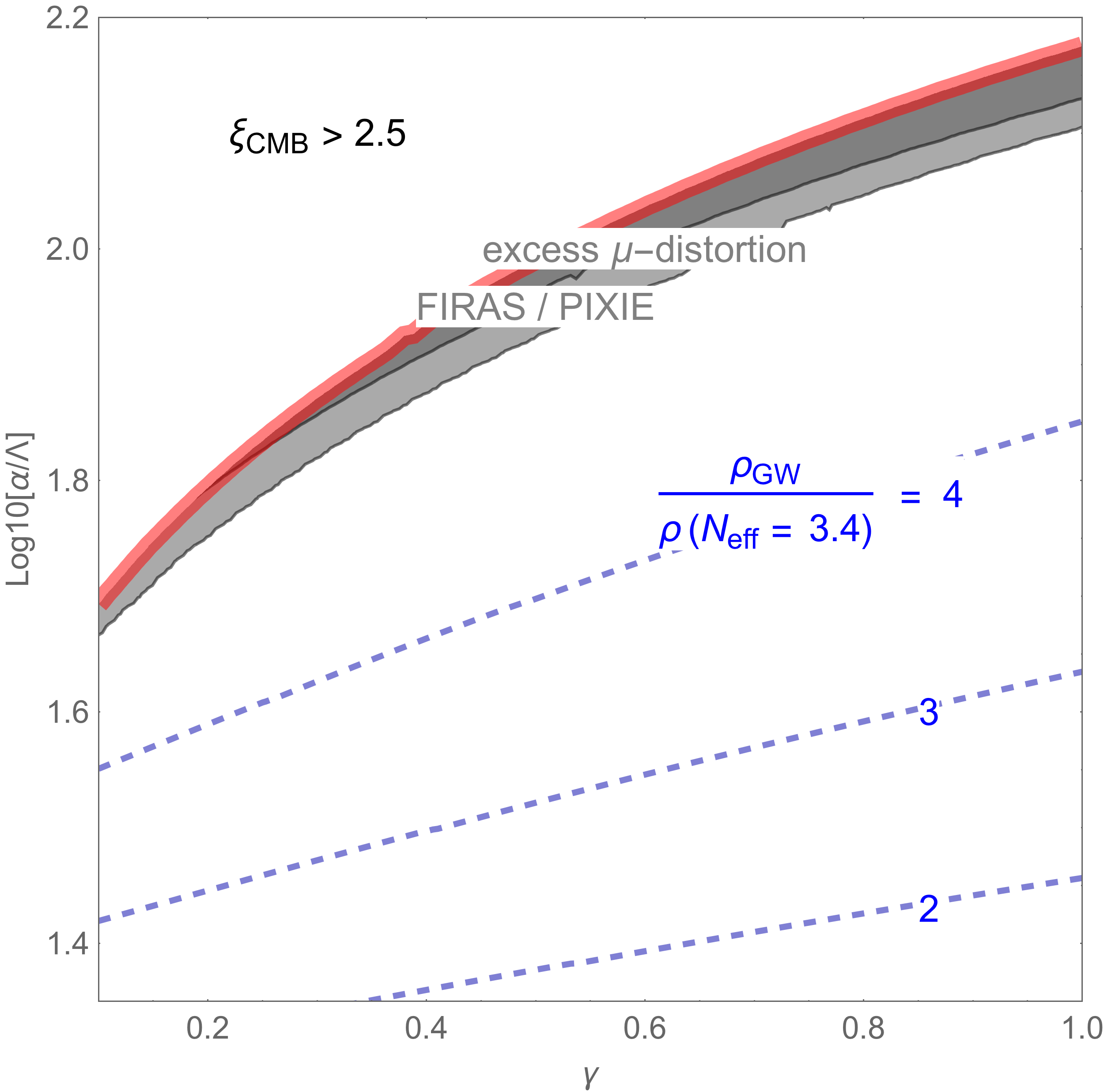}
\caption{$\mu$-distortion and contribution to $N_\text{eff}$ for the Starobinsky model. The dashed blue lines denote the factor by which $N_\text{eff}$ exceeds the current 95$\%$ CL bound of~\cite{Ade:2015xua} (see however comments in the main text), the dark grey coloured regions indicates the level of $\mu$-distortions excluded by FIRAS (COBE) at  95$\%$ CL. The light grey region shows the expeted sensitivity of PIXIE for an excess in the $\mu$-distortion above the vacuum contribution (based on the current $95\%$ CL region for $\Delta_s^2$ and $n_s$). The region on the top left is excluded by the non-gaussianity bound, $\xi_\text{CMB} < 2.5$.}
\label{fig:mudis}
\end{figure}

While the CMB is mainly sensitive to scales around $k \sim 10^{-2} \text{ Mpc}^{-1}$, a notable exception are distortions from a pure black body spectrum arising because the elastic and inelastic scattering processes of CMB photons freeze out at different times~\cite{Hu:1994bz,Pajer:2013oca,Meerburg:2012id}. These distortions are sensitive to the integrated scalar power spectrum in the range $50 \text{ Mpc}^{-1} \lesssim k \lesssim 10^4 \text{ Mpc}^{-1}$, corresponding to a frequency range of $10^{-15}~\text{ Hz} \lesssim f \lesssim 10^{-9}~\text{ Hz}$~\cite{Meerburg:2012id},
\begin{equation}
\mu \simeq 3 \int_{k_{D}(z_i)}^{k_{D}(z_f)} d\ln k  \; \Delta_s^2(k) \left[ e^{ - k^2/k_{D}^2(z)}\right]^{z_f}_{z_i} \,,
\end{equation}
with $k_D = 4 \times 10^{-6} z^{3/2} \text{ Mpc}^{-1}$ and $z_i = 2 \times 10^6$ ($z_f = 5 \times 10^4$)  denoting the redshift when the dominant inelastic (elastic) scattering processes for CMB photons freeze out. From Fig.~\ref{fig:DeltaS_all}, we note that only the Starobinsky model features an increase over the vacuum contribution in this frequency range. The bound from COBE / FIRAS constrains~\cite{Fixsen:1996nj} $\mu < 9 \times 10^{-5}$ at 95$\%$ CL and the most recent constraints from COBE / FIRAS plus TRIS~\cite{Zannoni:2008xx,Gervasi:2008eb} $\mu < 6 \times 10^{-5}$ impose a bound on the coupling parameter $\alpha$ comparable to the one from the limits on non-gaussianities in the CMB, $\xi_* < 2.5$, see Fig.~\ref{fig:mudis}. With a forcasted sensitivity of $\mu \lesssim 2 \times 10^{-8}$~\cite{Kogut:2011xw}, PIXIE is expected to reach the level of the vacuum contribution in this frequency range. In Fig.~\ref{fig:mudis}, the corresponding grey shaded region thus indicates the region in which the predictions for the Starobinsky model exceed the expected vacuum contribution for $f \leq 10^{-9}$~Hz. Note that $\mu$-type distortions constrain the spectrum when the gauge fields start to become important, but before the strong gauge field regime is reached. Compared to the bounds from primordial black holes and $N_\text{eff}$, this bound is thus less sensitive to the theoretical uncertainties inherent to this regime.

%% file: discussion.tex
The analysis above has revealed some universal features of pseudoscalar inflation as well as shed light on the existing parameter dependencies and degeneracies. As $\epsilon_\phi \rightarrow 1$ is a universal feature at the end of inflation, the asymptotic value for the GW spectrum at large frequencies does not depend on the underlying model of inflation and is insensitive to the variation of the coupling parameter $\alpha/\Lambda$ over a large range. Reducing the value of $\alpha/\Lambda$ however bans the increase in the GW spectrum to higher frequencies, moving it out of the range of conceivable detectors. A further remarkable feature is that low-scale inflation models, which feature a small tensor-to-scalar ratio, actually are more likely to yield a detectable GW signal in this setup due to an earlier and sharper rise of the GW spectrum - however in this case also the spectral index receives a stronger modification. Taking these two competing effects into account, the most promising model among discussed in this paper is the pseudoscalar Starobinsky model, allowing for a possible detection in ground-based and space-based interferometers as well as (marginally) in millisecond pulsar-timing arrays. This implies a remarkable complementarity between direct searches for GW and searches through CMB polarization for $r \gtrsim 10^{-3}$. Moreover, with the ongoing upgrades, LIGO/VIRGO is expected to reach a sensitivity to detect or rule out the $p=1$ and $p=2$ case in the next few years, if $\alpha/\Lambda$ is sizable. In the case of a positive detection, the upcoming eLISA mission would potentially allow to differentiate between these two cases, as well as constrain the value of $\alpha/\Lambda$.\\

\begin{figure}[ht]
\centering
\subfloat[][\emph{LIGO plot}.]
{\includegraphics[width=.45\columnwidth]{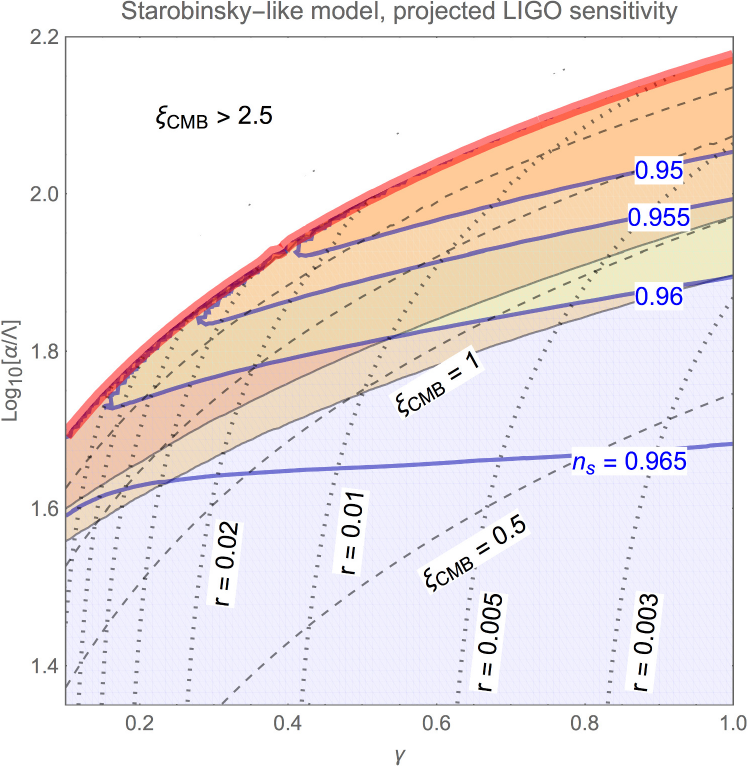}} \quad
\subfloat[][\emph{eLISA plot}.]
{\includegraphics[width=.45\columnwidth]{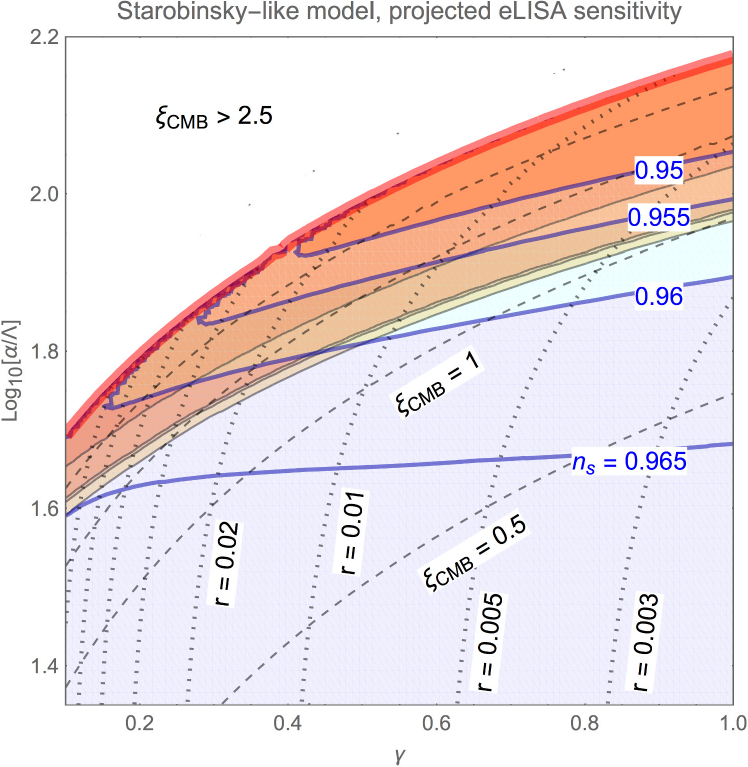}} 
\caption{Plot of the $(\alpha/\Lambda, \gamma)$ parameter space for the Starobinsky model with contour lines for $n_s$ (solid blue), $r = \{0.003, 0.005, 0.1, 0.2, 0.3, \dots \}$ (dotted) and $\xi_\text{CMB} = \{ 0.5, 1, 1.5, \dots \}$ (dashed). The orange shaded regions denote the projected sensitivity for advanced LIGO in the O2 and O5 run (left panel) and for eLISA in the C1 - C4 configurations (right panel).}
\label{fig:scan_plots}
\end{figure}

\noindent
The complementarity of CMB measurements and direct gravitational wave searches is made explicit in Fig.~\ref{fig:scan_plots} for the Starobinsky class of models ($p=2)$. In the parameter space spanned by $\alpha/\Lambda$ and $\gamma$, we numerically solve the equation of motion~\eqref{eq:approx_eq_motion_2} for the inflaton field, fixing $V_0$ (iteratively) to the value required by the COBE normalization. Fig.~\ref{fig:scan_plots} shows constraints from CMB measurements ($\xi_\text{CMB}$, $n_s$, $r$) as well as constraints and the projected sensitivity of direct gravitational wave detectors (eLISA and LIGO/VIRGO). The solid blue lines correspond to fixed values for $n_S$ with the shaded regions denoting the one and two sigma regimes; dotted lines correspond to fixed values for $r$ and dashed lines correspond to constant values for $\xi_{CMB}$. The upper bound $\xi < 2.5$ is marked by the red line. The orange shaded regions correspond to the observable regions for LIGO (left panel, evaluated at 50~Hz, runs O1, O2 and O5 as detailed in Sec.~\ref{sec:models}) and LISA (right panel, evaluated at 0.01~Hz, configurations C1 - C4 as detailed in Sec.~\ref{sec:models}). Remarkably, the current constraint on $\xi_\text{CMB}$ approximately coincides with the recently published data on LIGO run O1~\cite{TheLIGOScientific:2016wyq}. For $\gamma \gtrsim 0.2$, this moreover corresponds to the line in parameter space above which a too large spectral index is achieved as $N_2 \rightarrow N_\text{CMB}$. In summary, very different measurements are just starting to probe the viable parameter space of this model, and they are beginning to corner the parameter space from different directions: as searches for non gaussianities in the CMB and direct gravitational wave detection probe the region of large $\alpha/\Lambda$, searches for GWs in the CMB constrain the small $\gamma$ region.  A more precise measurement of $n_s$ could further narrow down the viable range for $\alpha/\Lambda$. Note that a similar analysis can be performed for the Hilltop models, with the corresponding parameter space spanned by $\alpha/\Lambda$ and $v$. However for $n_s > 0.9$, this class of models features an unobservably small GW signal in the sensitivity bands of eLISA and LIGO. Moreover, the tensor to scalar ratio is typically unobservably small and the spectral index lies below the observed value - generic features of models with $p \geq 3$.\\

\noindent
We now turn to the consequences for inflationary model building. A further universal feature of this setup is the reduction of the spectral index $n_s$. This may move inflationary models with a too flat (red) spectrum, such as e.g.\ supersymmetric hybrid inflation~\cite{Linde:1993cn,Binetruy:1996xj,Halyo:1996pp} with $n_s \simeq 0.98$, right into the sweet spot of the Planck CMB constraints. The observed reduction of $n_s$, parameterized by a reduction of the effective number of e-folds $N_* = N_\text{CMB} - \Delta N_*$ for a given inflation model, is degenerate with the uncertainties in the reheating phase which determine  $N_\text{CMB}$, the total number of e-folds of inflation. For a given inflation model, the parameter $\alpha$ hence allows to shift the predictions in the $n_s$ - $r$ - plane along the ``$N = 50 - 60$" lines depicted e.g.\ in Fig.~12 of \cite{Ade:2015lrj}.\\

\noindent
In our analysis of inflation models we have pursued a phenomenological approach, classifying inflation models by Eq.~\eqref{eq:Nparameterization} and studying in detail some well-known representative examples. A powerful tool on how to construct such models in a top-down approach has been put forward in Ref.~\cite{Linde:2012bt}. When constructing inflation models in supergravity, a common strategy to protect the flatness of the inflationary trajectory is to invoke a shift-symmetry~\cite{Kawasaki:2000yn}. Imposing this symmetry for the imaginary instead of the real part of the scalar component of the inflaton superfield supplies the desired pseudoscalar inflaton. A further concrete realization in supergravity, based on a superconformal symmetry, can be obtained from Ref.~\cite{Buchmuller:2013zfa} (see also~\cite{Kallosh:2013xya} for related work). Here the asymptotic behaviour of the Starobinsky potential is reproduced from a superconformal supergravity setup. Choosing $\chi$, the superconformal symmetry breaking parameter, to be positive flips the role of real and imaginary part of the complex scalar compared to the original setup of Ref.~\cite{Buchmuller:2013zfa}, yielding an interesting implementation of pseudoscalar Starobinsky inflaton.\\

\begin{figure}[ht]
\centering
\subfloat[][\emph{Scalar power spectra}.]
{\includegraphics[width=.44\columnwidth]{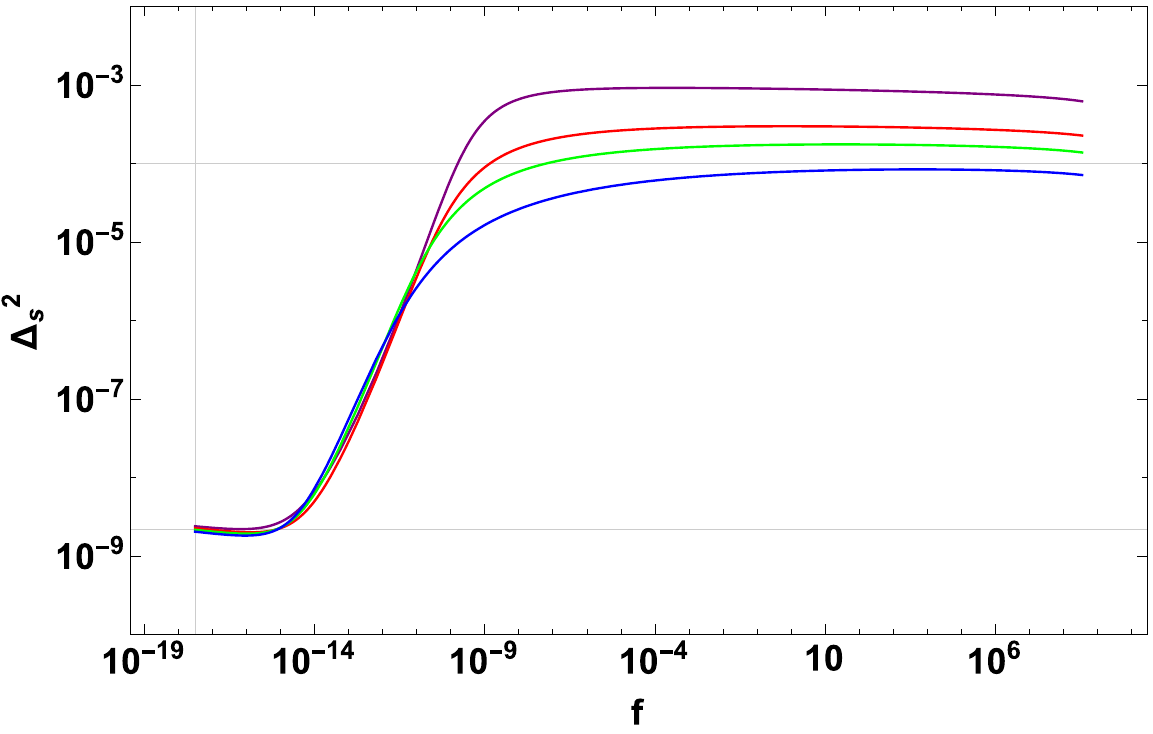}} \quad
\subfloat[][\emph{Tensor power spectra}.]
{\includegraphics[width=.5\columnwidth]{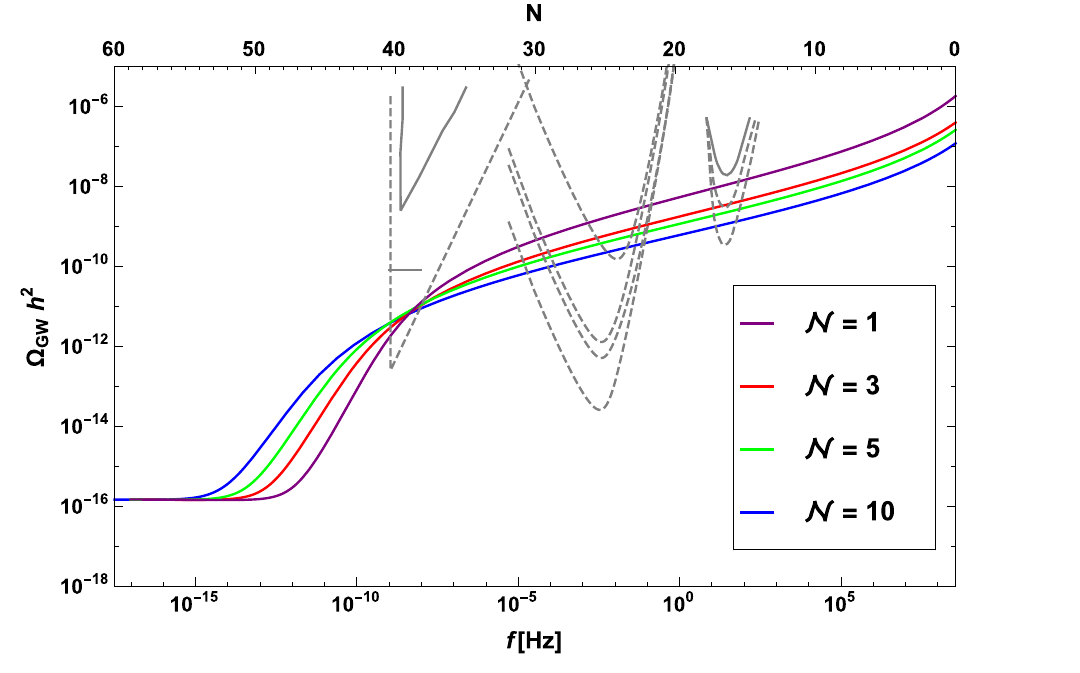}} 
\caption{Plot of scalar and tensor power spectra for Starobinsky models with $\alpha/\Lambda = 75, \  \gamma = 0.3, \ V_0 = 1.525 \cdot 10^{-9})$ for $\mathcal{N} = 1$ (purple), $\mathcal{N} = 3$ (purple), $\mathcal{N} = 5$ (red), $\mathcal{N} = 10$ (blue). The upper horizontal line in the plot on the left corresponds to the PBH bound and the lower one is the COBE normalization. In the plot for the tensor spectra we show the sensitivity curves for (from left to right): milli-second pulsar timing, eLISA, advanced LIGO. More details on these curves are given in Sec.~\ref{sec:explicit_models}.}
\label{fig:several_gauge_plots}
\end{figure}

\noindent
All the numerical plots that we have shown in this paper have been produced assuming $\mathcal{N} = 1$. However at inflationary energy scales several $U(1)$ gauge fields may be present. In this case the scalar and tensor power spectra are expected to be given by Eq.~\eqref{eq:scalar} and Eq.~\eqref{eq:OmegaGW} respectively. In Fig.~\ref{fig:several_gauge_plots}, we show the result of a direct numerical evaluation for the cases with $\mathcal{N} = 3,5,10$.
%
%
\noindent
As expected increasing the number of gauge fields affects the last part of the evolution without spoiling the dynamics at early times. In particular the scalar and tensor power spectra at CMB scales are exactly the same for all the models considered in the two plots. Moreover, the plot on the left shows that the estimate on the late time behavior for the scalar power spectra given in Eq.~\eqref{eq:scalar2} appears to be extremely accurate. As expected, the order one tension between the spectrum and the PBH bound is solved for the model with $\mathcal{N} = 10$. \\

\noindent
On the contrary, while looking at Eq.~\eqref{eq:OmegaGW} we would have naively expected an amplification of the GW spectrum when we increase $\mathcal{N}$, Fig.~\ref{fig:several_gauge_plots} clearly shows a different behavior. When we consider models with larger values of $\mathcal{N}$, in a first phase the spectra find a faster increase, but this lasts for a shorter period and the GW amplitude at later times is found to be suppressed. An explanation to this phenomenon may be be provided by reasoning on the way the gauge fields affect the dynamics. Several $U(1)$ will offer several channels for the decay of the inflaton, which will lead to a higher gauge field density and thus to the enhancement of the GW spectra. At the same time this will also accelerate the occurrence of the gauge field dominated regime, where the exponential growth of the GW spectrum is shut off - i.e.\ we find the parameter $\xi$, entering exponentially in Eq.~\eqref{eq:OmegaGW} to be surpressed (roughly $e^{4 \pi \xi}/\xi^8 \propto 1/{\cal N}^2$ from the equation of motion in the strong guage field regime), see Eqs.~\eqref{eq:eomC} to \eqref{eq:OmegaMax}. It is worth mentioning that this peculiar feature naturally provides a method to ease the tension with the $N_\text{eff}$ bound discussed in Sec.~\ref{sec:bounds}. In particular for models with $\mathcal{N} \gtrsim 5$ the present tension is completely removed. \\

\noindent
So far, our focus has been mainly on the very early epoch of cosmic inflation. But what is the fate of the produced gauge fields after the end of reheating? In the simplest case, the $U(1)$ gauge group is identified with SM hypercharge. In this case, the large abundance of gauge fields produced by the $\phi \tilde{F} F$  interaction during inflation as well as in the inflaton decay after the end of inflation will quickly populate the thermal bath~\cite{Barnaby:2011qe}. This suggests a very efficient reheating mechanism with an equation of state of $\omega \simeq 1/3$. Further implications of such a coupling to SM gauge groups are the presence of primordial magnetic fields and even a possible contribution to baryogenesis, see e.g.~\cite{Kusenko:2014uta, Anber:2015yca, Adshead:2015jza} for recent works. However it is also possible to imagine more complicated scenarios. Additional $U(1)$ symmetries may be present (and maybe even expected from the point of view of string theory at the high energy scales of cosmic inflation), broken spontaneously after the end of inflation.\footnote{In this case cosmic strings will be produced. Their non-observation in the CMB constrains the symmetry breaking scale to be around or below the GUT scale.} Depending on their couplings to the SM, the corresponding gauge bosons will decay into SM particles\footnote{An interesting example for such an additional $U(1)$ with couplings to the SM is the $U(1)_{B-L}$, with $B$-$L$ denoting the difference of baryon and lepton number, see~\cite{Buchmuller:2012wn,Buchmuller:2013dja} for possible further implications for early universe cosmology. } or into some hidden sector, contributing either to reheating or to dark matter.\\

\noindent
These details of the reheating have an important impact on the GW spectrum, which we have neglected so far. Eq.~\eqref{eq:OmegaGW} is based on some assumptions. Calculating the GW power spectrum today requires two ingredients: the power spectrum of the tensor fluctuations at the time of creation, i.e. when they exited the horizon during inflation, and the transfer function, which encodes the red-shift of the GW from horizon re-entry until today. The latter is given by:
\begin{equation}
T_k^2 \simeq \Omega_{R,0} \frac{g_*^k}{g_*^0} \left(\frac{g_{*,s}^0}{g_{*,s} ^k} \right)^{4/3} \,,
\label{eq:transferfunc}
\end{equation}
for modes $k$ re-entering during the radiation dominated regime. Here $g_*$ ($g_{*,s}$) counts the effective degrees of freedom entering the energy density (entropy) of the thermal bath. The superscript indicates evaluation at $t_k$ when the mode $k$ re-enters the horizon or today $(t_0)$, respectively.
For modes entering earlier, during the reheating phase, the transfer function depends on the respective equation of state. E.g. for a matter dominated reheating phase, there is a suppression factor of $(k_{RH}/k)^2$. Eq.~\eqref{eq:OmegaGW} hence assumes instantaneous reheating or a reheating phase with $\omega = 1/3$.\footnote{Possible changes in the degrees of freedom of the thermal bath, due to e.g.\ supersymmetry breaking, are also omitted in Eq.~\eqref{eq:OmegaGW}.} Analogous to the GW signal from the vacuum fluctuations during inflation, a deviation from this can lead to a suppression of the spectrum for frequencies larger than $f_\text{rh} \simeq 0.4~\text{Hz } (T_\text{RH}/10^7~\text{GeV})$, with $T_\text{RH}$ denoting the reheating temperature~\cite{Turner:1993vb,Seto:2003kc,Nakayama:2008ip,Buchmuller:2013lra}. For GUT-scale models of infation, such as Starobinsky inflation, this may hide a potential signal from the LIGO band, but typically not from the eLISA band located at lower frequencies.\\

\noindent
Recently, Ref.~\cite{Ferreira:2015omg} raised the question of the possible breakdown of a perturbative analysis for large values of $\xi$ as we consider in this paper towards the end of inflation. To clarify this point, we stress that while we assume perturbativity in the inflaton and tensor fluctuations, the gauge field production is an inherently non-perturbative process. We do not attempt a perturbative analysis of the gauge field, but work with the classical, non-perturbative background solution.\footnote{In fact, recent work~\cite{Peloso:2016gqs} finds perturbative control as long as $\xi \lesssim 4.7$, i.e.\ well into the regime where the GWs sourced by the gauge fields dominate over the vacuum contribution and produce an observable signal.} In particular, the requirement of perturbativity of the inflaton fluctuations imposes: 
\begin{equation}
\delta \phi \lesssim \Lambda/\alpha \,.
\end{equation} 
Throughout most of the evolution of the inflaton field, this is easily fulfilled. Note however that towards the end of inflation, inserting the asymptotic behaviour for the scalar power spectrum yields:
\begin{equation}
\label{eq:perturbativity_bound}
\langle \delta \phi^2 \rangle  \simeq \frac{\dot \phi^2}{H^2} \Delta_s^2 \simeq \phi_{,N}^2 \frac{1}{(2 \pi \xi)^2} =  \left(\frac{\Lambda}{\alpha \pi}\right)^2 \,. 
\end{equation}
Hence perturbativity is merely ensured by a factor of $1/\pi$, implying a potentially significant theoretical uncertainty in the asymptotic value of the scalar power spectrum\footnote{Notice that to obtain Eq.~\eqref{eq:perturbativity_bound} we use \cite{Linde:2012bt}:
\begin{equation}
\langle \zeta(x)^2 \rangle \simeq \mathcal{O}(1) \Delta_s^2(k) \ .	
\end{equation} A similar analysis can be carried out for tensor fluctuations. With $\langle h (x)^2 \rangle  \simeq C \Delta_t^2(k)$, where $C$ is a constant factor, and using Eq.~\eqref{eq:OmegaMax}, we can show that perturbativity is ensured for $C \lesssim 10^{5}$ .}. This is in particular relevant in view of the PBH bound in this regime. \\

\noindent Predictions in the  high-frequency regime are affected by several theoretical uncertainties. As bounds on the experimental side are improving rapidly (in particular direct GW detection through interferometers and improved $N_\text{eff}$ measurements through the next generation of CMB experiments), quantifying and improving on the theory uncertainties becomes crucial. Collecting some of the points previously mentioned, these uncertainties include: (i) A full quantum treatment of the perturbations in the strong backreaction regime, compared to the classical treatment of the background gauge field performed here. This most likely requires a lattice study of the non-perturbative system. (ii) Incorporation of the transfer function~\eqref{eq:transferfunc}, which may modify the spectrum at $k > k_{\text{RH}}$ depending on the equation of state during reheating. (iii) Possible decay of the energy stored in the gauge field into any particles $X$ charged under the corresponding gauge group, thus depleting the energy in the gauge sector as soon as $\langle \vec E^2 + \vec B^2 \rangle > m_X^2$,\footnote{We thank Cliff Burgess for pointing out this possibility.} see also~\cite{Kobayashi:2014zza}. \\

\noindent {An obvious extension of the framework  discussed in this paper is to consider the coupling of the pseudoscalar inflaton to non-abelian gauge groups. Initially, as long as the amplitude of the gauge field modes is small, the system will behave as in the Abelian case. However, as the exponential growth sets in, the non-abelian nature becomes important. Similar situations have been studied in lattice simulations for explosive gauge field production through preheating, both for the case of parametric resonance~\cite{Enqvist:2015sua} and a tachyonic instability~\cite{Skullerud:2003ki}, finding that the non-abelian interaction terms lead to a redistribution of the mode population towards higher values of $k$. In addition, effective mass terms may shut of the tachyonic instability prematurely.  These arguments indicate that the GW production should be less efficient in the non-abelian case. Similar questions have been adressed in the setup of so called chromo-natural inflation~\cite{Adshead:2012kp}. In this case, a coupling of a pseudoscalar inflaton to non-abelian gauge fields with a non-vanishing homogeneous vacuum expectation value can lead to a similar production of a chiral gravitational wave background, see e.g.~\cite{Dimastrogiovanni:2012ew,Obata:2016tmo}.  However since the simple estimates of Sec.~\ref{sec:review} no longer apply, a quantitative analysis of different inflation models coupled to non-abelian gauge fields is beyond the scope of the current paper.}

%% file: conclusions.tex
In this paper we have presented an updated discussion of a pseudoscalar inflaton non-minimally coupled with gauge fields. As widely discussed in literature, the  resulting generic production of gauge field quanta during inflation sensitively affects the scalar and GW spectra. In particular, this system features a tachyonic instability that leads to an exponential enhancement in the spectra as the inflaton speed increases towards the end of inflation. As a result, if the intensification is sufficiently strong, GWs produced with this mechanism can be observed with GW detectors such as LISA and advanced LIGO. Further striking observational signatures include the reduction of the spectral index $n_s$, the enhancement of the tensor-to-scalar ratio $r$ and the generation of non gaussianities in the CMB. \\

\noindent
Our analysis clarifies the parameter dependencies of these predictions due to the underlying inflation model. Classifying inflation models according to Ref.~\cite{Mukhanov:2013tua}, cf.\ Eq.~\eqref{eq:Nparameterization}, we quantify the effect of considering different universality classes of inflation (labeled by the parameter $p$) as well as varying the parameters within a given class. This study has both been performed analytically in Sec.~\ref{sec:analytical} and through numerical calculations, whose results are shown in Sec.~\ref{sec:models}. From the point of view of potential experimental signatures, we find that $p = 2$ (Starobinsky inflation) is the most promising candidate - a model which has recently received a lot of attention as it lies just in the sweet plot of the $n_s/r$ region preferred by Planck. As we show in Fig.~\ref{fig:scan_plots}, this model may lead to detectable chiral GW signals in both advanced LIGO and eLISA, with the parameter space further narrowed down by future CMB missions constraining $r$ and $n_s$. For a sizable value of  the inflaton gauge field coupling $\alpha/\Lambda$, this model is a fascinating candidate for multi messenger and multi frequency signals of cosmic inflation.\\

\noindent
In this paper we have performed a comprehensive discussion of the possibility of introducing a non-minimal coupling proportional to $\phi \tilde{F} F$ between a pseudoscalar inflaton and some gauge fields. Embedding this into a complete early time cosmological scenario confronts this setup with a number of constraints, as detailed in Secs~\ref{sec:models}, \ref{sec:bounds} and \ref{sec:discussion}. In this paper, our main focus is on consequences related to cosmic inflation, involving a broad range of CMB observables and GW signatures. In particular our analytical results may serve as guidelines for further inflation model building in this framework. Furthermore, it would be interesting to study in more detail the subsequent reheating phase as well as a possible connection to baryogenesis. In this light, not only connections to other eras of the early universe but also the concrete realization within particle physics models, identifying in particular the possible nature of the gauge group(s), poses interesting questions for future work.

%% file: appendix.tex
\begin{figure}[ht]
\centering
\subfloat[][\emph{Quadratic model with $\alpha/\Lambda = 35$ and $V_0~=~1.418~\cdot 10^{-11}$}.]
{\includegraphics[width=.45\columnwidth]{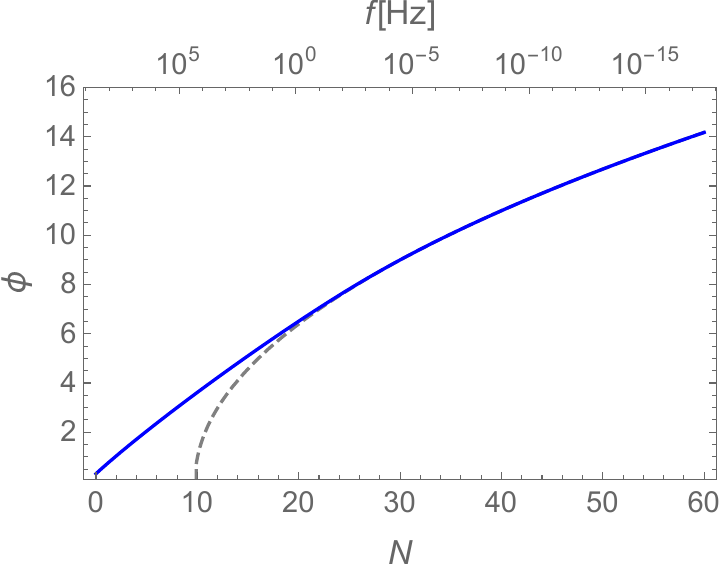}} \quad
\subfloat[][\emph{Starobinsky model with $\alpha /\Lambda = 75 $, $\gamma~=~0.3$, $V_0 = 1.525 \cdot 10^{-9}$}.]
{\includegraphics[width=.45\columnwidth]{PhiN_starobinsky}} \\
\subfloat[][\emph{Hilltop model with $q =4$, $\alpha/\Lambda =2000$, $v =0.1$ and $V_0 = 1.0 \cdot 10^{-21}$ }.]
{ \includegraphics[width=.48\columnwidth]{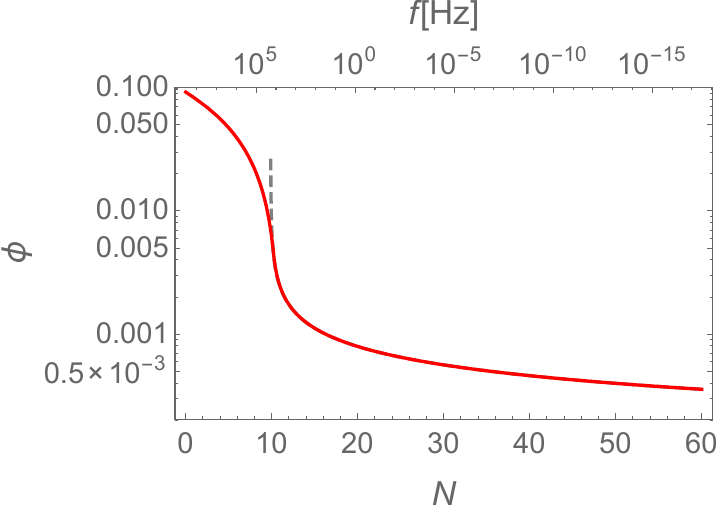}} \quad
\subfloat[][\emph{Hilltop model with $q =3$, $\alpha/\Lambda =2000$, $v =0.1$ and $V_0 = 3.6 \cdot 10^{-18}$ }.]
{ \includegraphics[width=.45\columnwidth]{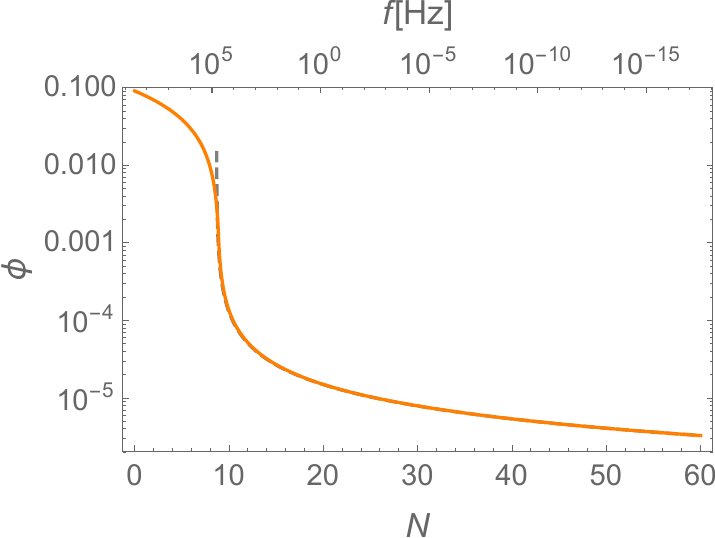}}
\caption{Evolution of inflaton field $\phi$ as a function of $N$ with (solid line) and without (dashed line) the non-minimal interaction with the gauge fields.}
\label{fig:phi_comparison}
\end{figure}

In this appendix we present the analysis of the models of Sec.~\ref{sec:models}. In particular we use the estimates of Sec.~\ref{sec:analytical} to explain the particular choice for the parameters used in the numerical simulations. The plots for $\xi$, $\Delta_s^2$ and $\Omega_{GW}$ are shown in Sec.~\ref{sec:models}, in Fig.~\ref{fig:xi_all}, Fig.~\ref{fig:DeltaS_all} and Fig.~\ref{fig:GW_all} respectively. The plots for the evolution of $\phi$ as a function of $N$ for all the models discussed in this paper is shown in figure Fig.~\ref{fig:phi_comparison}. Notice that for Starobinsky and Chaotic model $\phi$ decreases during inflation while in the case of Hilltop inflation with $p =3,4$ the field $\phi$ increases during the evolution. 

\subsection{Chaotic inflation, $p = 1$.}
\label{sec:chaotic}

As a representative for this class let us consider the case of Quadratic Chaotic Inflation:
\begin{equation}
\label{eq:models_quadratic}
V (\phi) = \frac{1}{2} \mu^2 \phi^2.
\end{equation}
Comparing Eq.~\eqref{eq:models_epsilon_general} with Eq.~\eqref{eq:epsilon_HV} we can easily find that for this model we have $\beta_p =1/2 $. The slow-roll parameters are then given by:
\begin{equation}
\label{eq:models_quadratic_slowroll}
\epsilon_V = \frac{1}{2N} \qquad \qquad \eta_V = \frac{1}{2N}
\end{equation}
The Planck constraint of Eq.~\eqref{eq:models_PLANCK} can be used to impose $N_* \simeq 60$. We can then use Eq.~\eqref{eq:epsilon_HV} to express the COBE normalization of Eq.~\eqref{eq:models_COBE} as $\mu \sim 6.106 \cdot 10^{-6} M_P$. Finally we can impose the constraint of Eq.~\eqref{eq:models_NG} to get $\alpha/\Lambda \lesssim 39$. \\

\noindent
Let us consider the evolution of a model with $\alpha/\Lambda \simeq 35$ and $V_0 = \mu^2/2 \simeq 1.418 \cdot 10^{-11}$.  The full numerical results  for this model is show in the plots of Figs. \ref{fig:phi_comparison}, \ref{fig:xi_all}, \ref{fig:DeltaS_all} and \ref{fig:GW_all}. The corresponding resulting values for the CMB observables are $n_s = 0.965, \ r  = 0.16$ and $\alpha_s = 0.000387$.
 The value of $\xi_1$ can be estimate by substituting $V_0$ into Eq.~\eqref{eq:analytical_xi1}:
\begin{equation}
 	\label{eq:models_xi1_chaotic}
 	\xi_1^6 = 2.03 \cdot 10^{-18} \cdot \exp\left( 4 \pi \xi_1\right), \qquad \longrightarrow \qquad  \xi_1 \simeq 3.89.
 \end{equation} 
To estimate the value of $\xi_{\text{max}}$ we start by estimating the value of $V(\phi)$ at the end of inflation. For this purpose we use $| V_{,\phi}/V| \simeq \sqrt{2}$ that gives $\phi_{\text{max}} \sim \sqrt{2}$. This directly implies $V(\phi_{\text{max}}) \simeq 2 V_0$. We can then use Eq.~\eqref{eq:xi_end} to get:
\begin{equation}
\xi_{\text{max}}^{3} \simeq 0.756 \cdot 10^{-15} e^{2 \pi \xi_{\text{max}}}, \qquad \longrightarrow \qquad \xi_{\text{max}} \simeq 6.43.
\label{eq:models_ximax_chaotic}
\end{equation}
The value of $\xi_2$ can be estimated from Eq.~\eqref{eq:knee_equation}. This requires an approximate expression for $V(\phi)$ as a function of $\xi_2$. For this purpose we can use Eq.~\eqref{eq:epsilon_HV} to get:
\begin{equation}
\xi_2^{7} \simeq 2.84 \cdot 10^{-10} e^{2 \pi \xi_2}, \qquad \longrightarrow \qquad \xi_2 \simeq 5.37, \qquad \phi_2 \simeq 6.52.
\label{eq:models_xi2_chaotic}
\end{equation}
Finally the estimated value for $\xi_{*}$ is given by Eq.~\eqref{eq:epsilon_HV}. The estimated values for $N_2$, and $N_1$ and $\Delta N_*$ are obtained by using Eq.~\eqref{eq:N_estimate}, Eq.~\eqref{eq:xigrowth2} and Eq.~\eqref{eq:N2}: $N_2 \simeq 19.33$, $\Delta N_* \simeq 8.71$, $N_1 \simeq 29.89$. The corresponding frequencies are given by Eq.~\eqref{eq:Nf}:
\begin{equation}
	f_1 \simeq 3.77 \cdot 10^{-5}~\text{ Hz}, \qquad \qquad f_2 \simeq 1.46~\text{ Hz} .
\end{equation}
As it is possible to see from Fig.~\ref{fig:GW_all}, these estimates are approximatively matching the values for these two frequencies in the full numerical computation.

\subsection{Hilltop inflation, $ 2 < p$.}
Let consider two different examples of models of this class. We begin by considering the potential of Eq.~\eqref{eq:models_hilltop_potentials} with $q = 4 $ and then we move to the case $q = 3$.

\subsubsection{Hilltop inflation, $p = 3$.}
\label{sec:models_hilltop_quartic}
Let us consider the potential:
\begin{equation}
\label{eq:models_hilltop_quartic}
V(\phi) =  V_0 \left[1 - \left (\frac{\phi}{v} \right)^4 \right]^2,
\end{equation}
Again we compare Eq.~\eqref{eq:models_epsilon_general} with Eq.~\eqref{eq:epsilon_HV} to find $\beta_p = v^4/128$. In this case the lowest order expression for the potential slow-roll parameters then read:
 \begin{equation}
\epsilon_V \simeq \frac{v^4}{128 N^3} , \qquad \qquad \eta_V \simeq - \frac{3}{2 N} .
 \end{equation}
 The constraints on the values for $n_s$ and $r$ then gives $N_* = 50$. We proceed by imposing $\xi_* \lesssim 2.5$ to get:
 \begin{equation}
  \frac{\alpha}{\Lambda} \lesssim 1.4 \cdot  \frac{10^4}{ v^2} .
 \end{equation}
Finally we can use the COBE normalization to determine $V_0$:
 \begin{equation}
  V_0 \simeq 3.26 \cdot v^4 \cdot  10^{-14}.
 \end{equation}
Notice that similarly to the case of Sec.~\ref{sec:starobinsky}, we are left with some free parameters. In this case we can choose an appropriate value for $v$ and $\alpha/\Lambda$ to satisfy all of the conditions.\\

\noindent
Let us consider $\alpha/\Lambda =2000$, $v =0.1$ and $V_0 \simeq 3.6 \cdot 10^{-18}$. This particular set of parameters gives $n_s =  0.940, \ r  = 9.93 \cdot 10^{-11}$ and $\alpha_s = 0.0039$. We proceed by following the procedure described in Sec.~\ref{sec:chaotic}: 
\begin{equation}
 	\label{eq:models_xi_hilltop4}
   \xi_1 \simeq 5.24, \qquad \qquad \xi_2 \simeq 7.34, \qquad \qquad \xi_{\text{max}} \simeq 9.13. 
 \end{equation} 
The estimated values for $N_2$, $\Delta N_*$ and $N_1$ are:
\begin{equation}
 	\label{eq:N_estimates_hilltop4}
   N_2 \simeq 11.59, \qquad \qquad \Delta N_* \simeq 11.29, \qquad \qquad N_1 \simeq 12.58. 
 \end{equation} 
The corresponding frequencies are given by Eq.~\eqref{eq:Nf}:
\begin{equation}
	f_1 \simeq 1.24 \cdot 10^{3}~\text{ Hz}, \qquad \qquad f_2 \simeq 3.34 \cdot 10^{3}~\text{ Hz} .
\end{equation}
As it is possible to see from Fig.~\ref{fig:GW_all}, $f_1$ and $f_2$ are approximatively matching the numerically obtained values. We should stress that for this value for $v$, it is not possible to choose $\alpha$ to saturate the CMB bound on $\xi$ without obtaining a huge value for $N_2$ that would be in contrast with Planck's constraints. As a consequence, the hilltop model does not yield a significant increase in $\Omega_{GW}$ compared to the Starobinsky model.

\subsubsection{Hilltop inflation, $p = 4$.}
We conclude this section by considering the cubic Hilltop potential:
\begin{equation}
\label{eq:models_hilltop_cubic}
V(\phi) =  V_0 \left[1 - \left (\frac{\phi}{v} \right)^3 \right]^2.
\end{equation}
In this case we have $\beta_p = v^6/72$. This directly leads to:
\begin{equation}
\epsilon_V \simeq \frac{v^6}{72 N^4} , \qquad \qquad \eta_V \simeq -\frac{2}{N}
\end{equation}
 The constraints on the values for $n_s$ and $r$ then gives $N_* = 60$. We proceed by imposing the small non gaussianity constraint i.e. $\xi_* \lesssim 2.5$ to get:
 \begin{equation}
  \frac{\alpha}{\Lambda} \lesssim 1.1 \cdot \frac{10^5}{ v^3} .
 \end{equation}
  We can then impose the COBE normalization:
 \begin{equation}
  V_0 \simeq 5.58 \cdot v^6 \cdot 10^{-16}.
 \end{equation}
As in sec \ref{sec:models_hilltop_quartic} we are then left with $v$ and $\alpha/\Lambda$ as free parameters. \\

\noindent
Considering the case with $\alpha/\Lambda =2000$, $v =0.1$ and $V_0 \simeq 1.0 \cdot 10^{-21}$. The corresponding values for $n_s, \ r$, and $\alpha_s$ are $n_s = 0.918, \ r = 3.2 \cdot 10^{-18}$ and $\alpha_s = 0.0017$. Again we proceed by following the procedure described in Sec.~\ref{sec:chaotic}: 
\begin{equation}
 	\label{eq:models_xi_hilltop3}
   \xi_1 \simeq 4.58, \qquad \qquad \xi_2 \simeq 8.79, \qquad \qquad \xi_{\text{max}} \simeq 10.5 . 
 \end{equation} 
The estimated values for $N_2$, $\Delta N_*$ and $N_1$ are:
\begin{equation}
 	\label{eq:N_estimates_hilltop3}
   N_2 \simeq 9.12, \qquad \qquad \Delta N_* \simeq 9.11, \qquad \qquad N_1 \simeq 9.14 . 
 \end{equation} 
The corresponding frequencies are given by Eq.~\eqref{eq:Nf}:
\begin{equation}
	f_1 \simeq 3.88 \cdot 10^{4}~\text{ Hz}, \qquad \qquad f_2 \simeq 3.95 \cdot 10^{4}~\text{ Hz} .
\end{equation}
\noindent
Again the values of $f_1$ and $f_2$ are in agreement with the ones of Fig.~\ref{fig:GW_all}. Similarly to the case of Sec.~\ref{sec:models_hilltop_quartic}, it is not possible to fix a value of $\alpha$ that gives $\xi_* \simeq 2.5$ and again this model does not yield a significant increase in $\Omega_{GW}$ compared to the Starobinsky model.

%% file: refs.tex
\providecommand{\href}[2]{#2}\begingroup\raggedright